\documentclass[12pt]{article}

\usepackage{cite,latexsym,graphicx,epsf}

\newcommand{\bp}{{\bf p}}
\newcommand{\bmp}{{\bf |p|}} 
\newcommand{\bq}{{\bf q}}

\newcommand{\bk}{{\bf k}}
\newcommand{\bmk}{{\bf |k|}}

\newlength{\figwidth}
\newlength{\figheight}
\begin{document}

\begin{titlepage}

\begin{flushright}
{September 1999}
\end{flushright}

\vskip 0.5cm

\begin{center}
	     {\large{\bf INFRARED DIVERGENCE IN QED$_3$ AT FINITE
TEMPERATURE}}

\vskip 0.6cm

\renewcommand{\thefootnote}{\mbox{\alph{footnote}}}

{{ Dominic Lee 
\footnote{E-mail address: d.lee2@physics.oxford.ac.uk }, Georgios Metikas
\footnote{E-mail address: g.metikas1@physics.oxford.ac.uk}
}}\\
\vskip 0.3cm
{Department of Physics, Theoretical Physics,\\
University of Oxford, 1 Keble Road, Oxford OX1 3NP}
\date{}
\end{center}

\vskip 2cm

\abstract{We consider various ways of treating the infrared divergence 
which appears in the dynamically generated fermion mass, when the  
transverse part of the photon propagator in N flavour
QED$_3$ at finite temperature is included in the Matsubara formalism.
This divergence is likely to be an artefact of taking into account
only the leading
order term in the $1 \over N$ expansion when we calculate the photon
propagator and is handled here phenomenologically by means of an
infrared cutoff. 
Inserting both the longitudinal and the transverse part of the photon
propagator in the Schwinger-Dyson equation we find the dependence of the dynamically generated fermion mass on the temperature and the cutoff parameters.
It turns out that consistency with certain statistical physics arguments
imposes conditions on the cutoff parameters. For parameters in the allowed
range of values we find that the ratio $r=2*Mass(T=0)/critical\;  temperature$
is approximately 6, consistent with previous calculations which neglected
the transverse photon contribution.}

\end{titlepage}

\setcounter{footnote}{0}

\section{Introduction} 

At zero temperature, a considerable quantity of work has been done
\cite{Pisarski,AppelquistSpontaneous,AppelquistCritical,Webb,Atkinson,Pennington,Curtis,DagottoAComputer,DagottoChiral,Nash,AtkinsonDynamical,Maris,Kondo} and has been extended to finite temperature more recently \cite{DoreyQED,DoreyThree,AitchisonPhase,AitchisonEffect,AitchisonBeyond,DominicEffect}. Our interest here is to carry forward the finite temperature calculations by improving upon the approximations made, as described below. In particular,
 we shall attempt to include the transverse  photon modes, which have
  so far been neglected  in all the calculations done using the Matsubara formalism.

At finite temperature Euclidean QED$_3$ in the reducible representation, there are three significant parameters $T_c$, $N_c$  and $r$ which need to be calculated. $T_c$ and $N_c$ are respectively the temperature and the number of flavours above which chiral symmetry is restored and at which there is a phase transition from a superconducting phase to a normal phase. $r$ is defined as the ratio of twice the zero temperature fermion mass to the critical temperature, $T_c$. This ratio is characteristic of the mechanism which leads to superconductivity. In Dorey and Mavromatos \cite{DoreyQED,DoreyThree} a calculation was
done using the Schwinger-Dyson equations at finite temperature and the
Matsubara formalism. In the following we shall be using the term
photon to describe not the electromagnetic photon but the statistical
photon. The approximations made in \cite{DoreyQED,DoreyThree} are as follows:

(i) The full photon propagator $\Delta_{\mu \nu}$ was calculated to leading order in $\frac{1}{N}$ in Landau gauge. 
\begin{equation}
\Delta^{m}_{\mu \nu}( \bmp ,\beta ) = \frac{A_{\mu \nu }}{p^2 + \Pi^{m}_{A}( \bmp ,\beta ) } +  \frac{B_{\mu \nu }}{p^2 + \Pi^{m}_{B}( \bmp , \beta )}
\end{equation}

\noindent  where $A_{\mu \nu}$ and $B_{\mu \nu}$, the longitudinal and transverse projection operators, are given by
\begin{eqnarray}
&& A_{\mu \nu}=\left( \delta_{\mu0}-{p_{\mu}p_0 \over p^2} \right) \left( {p^2 \over \bp ^2 } \right) \left( \delta_{0\nu}-{p_0p_{\nu} \over p^2} \right) \nonumber \\
&& B_{\mu \nu}=\delta_{\mu i} \left( \delta_{ij}-{p_ip_j \over 
 {\bf p}^2 }\right) \delta_{j \nu}
\end{eqnarray}

\noindent and $\Pi^{m}_{A}$, $\Pi^{m}_{B}$ are related to the one loop photon self-energy as follows: $\Pi^{m}_{\mu \nu } = \Pi^{m}_{A} A_{\mu \nu} + \Pi^{m}_{B} B_{\mu \nu}$. 

(ii) Only the $\mu = \nu = 0$ part of $\Delta^{m}_{\mu \nu}$ was retained.

(iii) Only the $m=0$ ($p_0 = 0$) part of $\Delta^{m}_{0 0}$ was retained, retardation effects being neglected.

Approximations (ii) and (iii) will be referred to as the instantaneous exchange approximation which corresponds to considering the quantum statistical mechanics of an ensemble of Dirac fermions interacting via the static two-body potential instead of a full field-theoretical treatment.

(iv) The fermion mass $\Sigma $ was assumed to depend only on temperature and not on frequency or momentum. \[ \Sigma^{m}(\bmp , \beta ) \approx \Sigma^{0}(0, \beta ) = \Sigma(\beta ). \]

(v) the fermion wave function $A$ was neglected, $A=1$.

As a consequence of (iv), (v) and the Ward-Takahashi identity, the full vertex $\Gamma_{\mu}$ was taken to be $e \gamma_{\mu}$.

The results were that there is dynamically generated mass which is a decreasing function of the temperature $T$ and vanishes for temperatures above a critical temperature $T>T_c$, $T_c$ is itself a decreasing function of the number of flavours $N$. No critical number of flavours $N_c$ was found when the zero temperature limit was taken and no $N_c$ was expected at finite temperature. This is in agreement with the zero temperature calculation of \cite{Pisarski}, because this work adopts the constant mass approximation (iv) as well. The ratio $r$ was found to fluctuate around $10$ as $N$ increases. This is a large value compared to the BCS $r \simeq 3.54 $ and seems to be phenomenologically correct, as there are indications that the layered CuO superconductors indeed have larger $r$ than BCS systems. The origin of this phenomenon is traced to the appearance of a temperature dependent mass for the temporal photon (plasmon mass), which implies that external electric fields are screened by thermal excitations. 

 In Aitchison, Dorey, Klein-Kreisler and Mavromatos
\cite{AitchisonPhase}, the constant mass approximation (iv) was
relaxed and the momentum dependence was taken into account \[
\Sigma^{m}(\bmp , \beta ) \approx \Sigma^{0}(\bmp ,\beta). \] The
results were similar to those of  \cite{DoreyQED,DoreyThree}, the only
difference being that a temperature dependent critical number of
flavours $N_c(T)$ was found. This is the same behaviour seen at zero
temperature when momentum dependence was taken into account
\cite{AppelquistSpontaneous,AppelquistCritical}. The zero temperature
limit of the $N_c(T)$ found in \cite{AitchisonPhase} is just greater
than $2$ and therefore does not coincide with the value $3.2$ of
\cite{AppelquistSpontaneous,AppelquistCritical}. This is due to the
fact that the finite temperature calculation of \cite{AitchisonPhase}
employed the instantaneous exchange approximation which was not used
in the zero temperature calculations
\cite{AppelquistSpontaneous,AppelquistCritical} . It should be
stressed that the relaxation of (iv) did not affect the value of $r$
which remained around $10$.

Following the zero temperature arguments of \cite{Webb,Atkinson,Pennington,Curtis}, Aitchison and Klein-Kreisler \cite{AitchisonEffect} performed a finite temperature calculation dropping approximations (iv) and (v) simultaneously. The Pennington-Webb vertex $\Gamma_{\nu}=\gamma_{\nu} A $ of \cite{Webb,Atkinson} was adopted. Unlike the more elaborate vertex of \cite{Curtis} (Ball-Chiu vertex + a non-trivial transverse part), the Pennington-Webb vertex does not satisfy the Ward-Takahashi and Ward identities. However it yields results which are in qualitative agreement with those of \cite{Curtis}.
Although works \cite{Webb,Atkinson,Pennington,Curtis} show that taking into account the momentum dependence of the mass and the wave-function renormalisation leads to the conclusion that there is no $N_c$ at zero temperature, \cite{AitchisonEffect} suggests that this not the case at finite temperature. The reason can be traced to the different infrared behaviour of the photon self-energy at zero and at finite temperature. Thus the results of \cite{AitchisonEffect} are very similar to those of the earlier finite temperature work \cite{AitchisonPhase} and the value of the ratio $r$ was once again found to be approximately $10$.

The $r=10$ feature of QED$_{3}$ seemed to be remarkably robust, surviving all refinements, until Aitchison \cite{AitchisonBeyond} showed that dropping the instantaneous exchange approximation (ii), (iii) leads to a considerable reduction of the value of $r$ from $10$ to $6$. 
One obvious problem with the instantaneous approximation is that it cannot reproduce the well-studied zero temperature limit. This is because as $T \rightarrow 0$, all frequency components are to be included; but the instantaneous
approximation retains only the $n=0$ component of the photon propagator, so the solution cannot join continuously on to the $T=0$ solution. A second problem (indicated here for the first time) may be that the instantaneous approximation cannot reproduce the requirement that ${ d \Sigma \over dT} \rightarrow 0$ as $T \rightarrow 0$; we shall show in Section 5 that this requirement must be true of physical solutions, if the thermodynamic arguments of that section hold.

The first study of retardation effects in the S-D equation in QED$_3$
at finite temperature \cite{AitchisonBeyond}, was performed in the real time formalism. This has the advantage that there is no discrete sum over frequency components, so that the zero temperature result
could be recovered straightforwardly. The reduced value $6$ was calculated for $r$. No critical number of flavours $N_c$ was found. This calculation, however, used an approximate form for the photon propagator, which did not treat correctly the non-analytic structure near the origin. Therefore some doubt remains as regards the reliability of the results.

 Finally Lee in \cite{DominicEffect} made an attempt to drop the instantaneous exchange approximation (ii) and (iii) in the Matsubara formalism this time. Convenient  approximate forms for both the longitudinal and the transverse part of the full photon propagator were found  to leading order in $1/N$, having the correct behaviour for small $3$-momentum. Both the constant mass approximation and a second approximation, where the mass was taken to be frequency
dependent, were investigated. As expected, in the former case no $N_c$ was found whereas in the latter case there was a $N_c$. In both cases $r \simeq 6 $ in agreement with \cite{AitchisonBeyond}. However the problem with \cite{DominicEffect} is that, in the S-D equation for the full fermion propagator, the transverse part of the full photon propagator was discarded, because of its divergent infrared behaviour. This new approximation is qualitatively similar to (ii), since $\Delta^{m}_{00}$ has no transverse part as well. Thus this paper, although it claims to discard both (ii) and (iii), in reality drops only (iii) and substitutes (ii) with an equivalent approximation.

Our purpose is to go beyond \cite{DominicEffect} by including the transverse part of the full photon propagator in the S-D equation for the full fermion propagator at finite temperature. In other words we shall completely discard (ii) and (iii) but retain (i), (iv) and (v).

The infrared divergence of the transverse contribution of the full photon propagator makes the S-D equation meaningless and we must find some way of dealing with it. One might at first think that the trouble has arisen, because the photon self-energy has been calculated with a massless fermion in the loop, instead of the self-consistently generated mass $\Sigma$. However including $\Sigma$ in the loop calculation does not help, as we show in Appendix A. We are therefore forced to consider the introduction of some form of infrared cutoff, if we want to proceed at all. We shall consider two types of cutoff: the first is on the momentum integral of the S-D equation, which
we place at $\bmk = \epsilon k_B T$; the second is a mass cutoff
 in which we give the zeroth mode of the transverse photon
contribution a squared mass of the form $\delta \alpha \ln 2 k_B T$ similar to the plasmon mass of the temporal part of the photon propagator. We may view these cutoffs as a phenomenological way of taking into account terms beyond the leading order in $1/N$, since it seems likely that the infrared divergence would be removed, if the full (all orders in $1/N$) zero frequency contribution could be calculated. The introduction of infrared cutoffs of both the integral and the mass forms is a usual practice in the zero temperature literature, although there is no infrared divergence in this case \cite{KondoCutoff, AitchisonDeviations, AitchisonNontrivial}.  Our
purpose then, is to investigate how the fermion 
mass $\Sigma$ depends on the two types of cutoff and to attempt to extract sensible physical conclusions from these calculations.

In Section 2 we formulate the gap equation for $\Sigma$. In Section 3 we consider what we call ``extreme'' solutions  in which the infrared cutoffs $\epsilon$ and $\delta$ are taken to be so small that the infrared divergence is the dominant effect. In Section 4 we
 present numerical solutions for a range of values of $\epsilon$ and $\delta$. We find that there are two classes of solution: those at ``large'' $\epsilon$ and $\delta$ for which ${d \Sigma \over dT} \leq 0$ for all $T$  and those at ``small'' $\epsilon$
 and $\delta$ for which ${d \Sigma \over dT}>0$ at small $T$. In Section 5 we shall argue that the latter class is unphysical  by considering the entropy of a dilute gas of fermions with mass $\Sigma(T)$. Accepting this argument, it follows that there exists
 critical values $\epsilon_c$ and $\delta_c$ below which the solutions are unphysical, in the sense of ${d\Sigma \over dT} >0$ for small $T$. These critical values 
are obtained in Section 6 by finding a simple analytical form for the numerically calculated $\Sigma(T,\epsilon \mbox{ or } \delta)$, which may itself be useful in other contexts. Some final comments are made in Section 7.

\section{The Schwinger-Dyson equation for $\Sigma$} 

The Lagrangian of massless, Euclidean QED$_3$ with N fermion flavours is 

\begin{equation}
 L=-\frac{1}{4} f_{\mu \nu}f^{\mu \nu} + \sum_{i=1}^{N} \bar{\psi_i}
( i \!\not\!\partial  -e \!\not\! a  )\psi_i 
\label{QED3Lagrangian}
\end{equation}

\noindent We consider the reducible representation of the Dirac algebra so that
the above Lagrangian has two continuous chiral symmetries.
 In the Matsubara formalism, the full Schwinger-Dyson equation for the full fermion propagator at finite temperature takes the form 

\begin{equation}
S^{-1}_F(p_f)=S^{(0)-1}_F(p_f) - {e \over \beta} \sum_{n = - \infty }^{ \infty }
\int {d^2 \bk \over (2 \pi)^2 } \gamma_{\nu} S_F(k_f) \Delta_{\mu \nu}(q_b) \Gamma^{\mu}(k_f,p_f).
\label{SD}
\end{equation}

\noindent where we have used the subscripts $f$ and $b$ to denote fermionic and bosonic momenta respectively and

\begin{eqnarray*}
p_{f}=(p_0, \bp ) && p_0=\frac{(2n+1)\pi }{\beta} \\
k_{f}=(k_0, \bk ) && k_0=\frac{(2m+1)\pi }{\beta} \\
q_{b}=p_{f}-k_{f}=(q_0, \bq ) && q_0=\frac{2(n-m)\pi }{\beta}
\end{eqnarray*}

\noindent After applying approximations (i), (iv), (v), taking the spinor space trace and performing the trivial angular integration, (\ref{SD}) reads

\begin{equation}
\Sigma(\beta)=\frac{\alpha}{N \beta} \sum_{m=-\infty}^{\infty} \int_{0}^{\infty}  \frac{ \bmk d\bmk }{2 \pi} \Delta^{-m}{\mu \mu}(\bmk , \beta) \frac{\Sigma(\beta)}{ \left[ \frac{ (2m+1)\pi }{\beta } \right]^2 +\bk ^2 + \Sigma(\beta)^2} \label{sgapequation}
\end{equation}

\noindent where

\begin{equation}
\Delta^{-m}_{\mu \mu} (\bmk , \beta )= \frac{1}{ \left[ \frac{2m \pi}{\beta} \right]^2 +\bk ^2 + \Pi^{-m}_A (\bmk, \beta )} +  \frac{1}{ \left[ \frac{2m \pi}{\beta} \right]^2 +\bk ^2 + \Pi^{-m}_B (\bmk, \beta )}. 
\label{DeltaQED}
\end{equation}

\noindent A trivial solution of (\ref{sgapequation}) is that $\Sigma $ is identically zero for any temperature. However this solution for the fermion mass would rule out QED$_3$ as a model of superconductivity. If QED$_3$ is to be such a model, it should be possible to prove that there exists a critical temperature $T_c$ (see Appendix B) below which there is a non-zero solution of (\ref{sgapequation}). Thus we proceed by assuming that $\Sigma$ is not identically zero and seeking a self-consistent solution of the resulting equation:

\begin{equation}
1=\frac{\alpha}{N \beta} \sum_{m=-\infty}^{\infty} \int_{0}^{\infty}  \frac{ \bmk d\bmk }{2 \pi} \Delta^{-m}{\mu \mu}(\bmk , \beta) \frac{1}{ \left[ \frac{ (2m+1)\pi }{\beta } \right]^2 +\bk ^2 + \Sigma(\beta)^2} \label{gapequation}
\end{equation}

 Both (\ref{gapequation}) and (\ref{DeltaQED}) are well-known and appear 
for example in \cite{DoreyQED,DoreyThree}. The exact expressions for $\Pi^{-m}_A$ and $\Pi^{-m}_B$ are also given in \cite{DoreyQED}. However (\ref{gapequation}) is too hard to attempt even a numerical solution. This is exactly why a further approximation, such as the instantaneous exchange approximation, was necessary. Recently in \cite{DominicEffect} it was shown that a less crude approximation is sufficient to render (\ref{gapequation}) numerically solvable.

\begin{eqnarray}
\mathrm{For} \  m=0 && \Pi^{0}_{A} = \Pi^{0}_{3} \simeq  \frac{\alpha }{\beta }  \frac{1}{8} \left[ \bmk \beta + \frac{16 \ln{2} }{\pi} \  e^{ - \frac{\pi}{16 \ln{2} } \bmk \beta } \right] \nonumber \\
                 && \Pi^{0}_{B} = \Pi^{0}_{1} \simeq \frac{\alpha }{\beta }
\frac{\bmk \beta }{8} \left[ 1 - e^{ - \frac{\pi}{16 \ln{2}} \bmk \beta } \right]. \nonumber \\
\mathrm{For} \ m \neq 0 && \Pi^{-m}_{A} \simeq  \Pi^{-m}_{B} \simeq \frac{\alpha }{8} \sqrt{ \left[ \frac{2 m \pi }{\beta } \right]^2 + \bk ^2}.
\end{eqnarray}

\noindent As noted in \cite{DominicEffect}, when $ \beta \bmp \rightarrow 0 $, $ \Pi^{0}_{1} \rightarrow f \beta^2 \bmp ^2 \frac{\alpha}{\beta}$. $f$ is a constant which is found to be $\frac{\pi}{128 \ln{2}}$ in our approximation of the photon propagator; this differs by a factor of $3/4$ from the $\bmp \rightarrow 0$ behaviour of the exact result (see Appendix A). For $\Pi^{0}_{3}$ the low $\bmp $ behaviour is different; in this limit, $\Pi^{0}_{3} \rightarrow \frac{2 \ln{2} }{\pi} \ \frac{\alpha}{\beta}$, this constant value being the plasmon mass.
 
 As we have already explained, \cite{DominicEffect} does not make full use of these more accurate forms and in particular of the expressions for $\Pi^{0}_{B}$ and $\Pi^{m}_{B}, m  \neq 0$.
 The approximate form for $\Pi^{0}_{A}$ had also been employed in the earlier work \cite{AitchisonPhase} in combination with the instantaneous exchange approximation.

\noindent On rearrangement (\ref{SD}) reads

\begin{equation}
1= \frac{a}{ 2N \pi} [S_L(a,s)+S_T(a,s)].
\label{rearranged}
\end{equation}

\noindent $S_T(a,s)$ is the contribution from the transverse part and takes the form:

\begin{eqnarray}
S_T(a,s) &=& \int^{\infty}_{0} x \; dx \left[ { 1 \over x^2 + \beta^2 \Pi_1^0 } \right.
\; { 1 \over x^2 + \pi^2 + a^2s^2 } \nonumber \\
&\; +& \sum_{m=1}^{ \infty }  { 1 \over x^2 + (2 \pi m)^2 +
 0.125 a [x^2 + (2 \pi m )^2]^{1/2} }  \nonumber \\
&\; \times&  \left( { 1 \over x^2 + [2 \pi(m+1/2)]^2 +a^2s^2 }
\left. + { 1 \over x^2 + [2\pi (m-1/2)]^2 +a^2s^2 }  \right) \right] 
\label{transverse}
\end{eqnarray}

\noindent where $x=|{\bf k}| \beta$, $a=\alpha \beta$ and
$s=\Sigma/\alpha$. $S_L$ is the longitudinal contribution to (\ref{rearranged}) and is exactly the same as $S_T$ in (\ref{transverse}), except
 for $\Pi^0_1$ being replaced by $\Pi^0_3$.

 The low $|{\bf k}|$ behaviour of $\Pi^0_1$
is $\Pi^0_1 \propto {\bf k}^2$. 
It follows that there is an infrared divergence in (\ref{transverse}), which is absent in the case of $S_L$ due to the presence of the plasmon mass in $\Pi^{0}_{3}$.
We note that using the finite fermion mass self-energy, 
$\Pi^{\Sigma}_{\mu \nu}$ (Appendix A), instead of the zero mass quantity 
$\Pi_{\mu \nu}$, would not cure this divergence.
 We are left with one other alternative at leading order in $1/N$: to
impose an infrared cutoff.

We shall look at two types of cutoff. In the first, which we call the
integral cutoff, we replace the lower limit of the $|{\bf k}|$-integral in (\ref{gapequation}) by $\epsilon k_B T$, $\epsilon$ being an arbitrary constant.
 On rearrangement of the terms leading to
 (\ref{rearranged}) and (\ref{transverse}), this cutoff becomes the lower limit of the $x$-integral, $\epsilon$. The idea behind 
choosing such a cutoff is that at $T=0$ there is no infrared divergence in (\ref{rearranged}) and we shall want our $T=0$ solution to join smoothly on to our $T \neq 0$ solutions, when we consider the contribution from all the modes in $S_L$ and  $S_T$. The form  $\epsilon k_BT$ is the simplest cutoff on the
$|{\bf k}|$ integral which ensures this, for as $T \rightarrow 0$ the cutoff 
on the $|{\bf k}|$-integral tends smoothly to zero.

In the second type of cutoff we shall add an extra plasmon-like term
$\delta \alpha ln2 / \beta$ to  $\Pi^0_1$, where again $\delta$ is an arbitrary constant; we call this procedure the mass cutoff. Again at $T=0$ this will vanish leaving a smooth $T \rightarrow 0$ limit.

We are able to evaluate analytically all the $x$-integrals for the $m
\neq 0$ modes of (\ref{rearranged}).
 In our evaluation of the $x$-integrals in the sums, we use an
approximation: we neglect $\epsilon$ in  the integrals of all the modes except
the $m=0$ mode in $S_T$  and $S_L$. 
This approximation is only valid for $\epsilon < 2 \pi$, which is not a problem, as $\epsilon $ is an infrared cutoff.

Using this further approximation gives us the following expression in the integral cutoff case:
\begin{eqnarray}
 1= \frac{Q(a,s)}{2\pi N} \equiv \frac{a}{ 2\pi N} &\times & \left[ \sum^{\infty}_{m=1} 2I( 2 \pi m ,0.125a,( a^2s^2 + [2 \pi (m+1/2)]^2 )^{1/2}) \right. \nonumber \\
&+& \sum_{m=1}^{\infty} 2I( 2 \pi m ,0.125a,( a^2s^2+[2 \pi (m-1/2)]^2)^{1/2})  \nonumber \\
&+& \left. \int_{\epsilon}^{\infty} \frac{x \; dx} {x^2+\pi^2+a^2s^2} \left( \frac{1}{ x^2+ \Pi_3^0(x) \beta^2 } + \frac{1}{ x^2+ \Pi_1^0(x) \beta^2} \right) \right] 
\label{integralQED}
\end{eqnarray}

\noindent where

\begin{equation}
I(d,a,c) = {1 \over 2(a^2+c^2-d^2)} \ln \left[ {c^2 \over (d+a)^2} \right] + {a \over (c^2-d^2)^{1/2}(a^2+c^2-d^2) } \arctan \left[ (c^2-d^2)^{1/2} \over d \right].
\end{equation}

\noindent For the mass cutoff our equation is similar but with the zeroth mode contribution in (\ref{integralQED}) being replaced by:

\begin{equation}
 \int_{0}^{\infty}{x \; dx \over x^2+\pi^2+a^2s^2 } \left( {1 \over
x^2+ \Pi_3^0(x)\beta^2}+{1 \over x^2+ \Pi_1^0(x)\beta^2+a \delta \ln
2} \right). 
\label{mass}
\end{equation}

\noindent Although we have argued that terms of higher order in $1/N$
would regulate the infrared divergence, we have no way of telling whether the mass or integral cutoff is the best way of approximating such terms
 (although we shall see later that of these two cutoffs, if QED$_3$ is to be a model of superconductivity, the integral cutoff is preferred)
 or what values of $\epsilon$ or $\delta$ give the best approximation without explicitly calculating these terms.
 We can however start by considering the limiting
 case of very small $\epsilon$ or $\delta$, where we can make some progress analytically. 
We shall find in Section 3 what appear to be clearly unphysical features of the solution,
 thus confirming that the divergence is physically serious and suggesting that
 $\epsilon$ (or $\delta$ ) cannot be too small for consistent physical results.
 This will lead us to a numerical exploration of the solutions for various $\epsilon$($\delta$) in Section 4. We shall find that for $\epsilon$($\delta$) less than a certain
 critical value $\epsilon_c$ $(\delta_c)$ the quantity ${d \Sigma \over dT} \geq 0$ as $T \rightarrow 0$, which is physically undesirable, as we shall explain by a thermodynamic argument in Section 5. Requiring ${d \Sigma \over dT} \leq 0$ for all $T$  therefore imposes
 a constraint on the possible smallness of $\epsilon(\delta)$ and hence (on our interpretation) of the higher order contribution
 in $1/N$. We shall also consider how sensitive the ratio $r= {2\Sigma(0) \over k_BT_c}$ is to variations of $\epsilon(\delta)$ in
 the allowed regions $\epsilon > \epsilon_c$, $\delta>\delta_c$.

\section{The extreme solutions for $\Sigma$ }

In this section we shall be interested in the behaviour of $\Sigma$ for very small $\delta$ and $\epsilon$ for which we can see exactly what behaviour the infrared divergence induces in $\Sigma$.

To find the solution of (\ref{integralQED}) at extremely small $\epsilon$($\delta$) we need only retain the zeroth transverse mode;
 at sufficiently small $\epsilon$($\delta$) this dominates (\ref{integralQED}) (at large enough $Tk_B/\alpha$) so that all other terms may be neglected.
 We are therefore left with an equation - in the case of an integral
cutoff - of the form 
\begin{equation}
1={a \over 2\pi N} \int^{\infty}_{\epsilon} {x \over x^2 + \Pi^0_1(x)\beta^2}{dx \over x^2+\pi^2+a^2s^2}.
\label{zerothtransverse}
\end{equation}

\noindent Now we know that for small $x$, $\Pi^0_1(x) \beta^2$ goes as
$afx^2$. Since we are interested only in the singular contribution, 
 when $\epsilon \rightarrow 0$, we may impose an ultraviolet cutoff $\Lambda$. The purpose of $\Lambda$ is merely to simplify the analysis: below $
\Lambda$ we shall, to a good approximation, be able to replace $\Pi^0_1$ with $afx^2$; above $\Lambda$, we shall be able to neglect
 contributions from the various momenta, due to the fact that, for
extremely small $\epsilon$ (and $\delta$), the integral from $\Lambda$ to $\infty$ is negligible compared to the infrared divergent integral from $\epsilon$ to $\Lambda$. By replacing $\infty$ by $\Lambda$ we are able to write (\ref{zerothtransverse}) as:

\begin{equation}
1={a \over 2\pi N} \ {1 \over 1+fa}  \int^{\Lambda}_{\epsilon} {dx \over x(x^2+\pi^2+a^2s^2) }.
\label{uvzerothtransverse}
\end{equation}

\noindent The integral in (\ref{uvzerothtransverse}) can be easily evaluated, yielding the following equation for $s$:
\begin{equation}
1= \frac{a}{2\pi N}\  \frac{1}{1+fa}\  \frac{1}{ \pi^2 + a^2 s^2} 
\left[ \ln{ \frac{ \Lambda }{ \epsilon} } - \ln{ \sqrt{ \frac{\Lambda^2 + \pi^2 + a^2 s^2}{ \epsilon^2 + \pi^2 + a^2 s^2 } }} \right].
\label{extremeintegral}
\end{equation}

\noindent We note that, since $\Lambda$ is the ultraviolet and $\epsilon$ the infrared cutoff, \[ \Lambda \gg \epsilon \Leftrightarrow \frac{\Lambda }{\epsilon } \gg \sqrt{ \frac{\Lambda^2 + \pi^2 + a^2 s^2}{ \epsilon^2 + \pi^2 + a^2 s^2 } } \Leftrightarrow  \ln{ \frac{ \Lambda }{ \epsilon} } \gg  \ln{ \sqrt{ \frac{\Lambda^2 + \pi^2 + a^2 s^2}{ \epsilon^2 + \pi^2 + a^2 s^2 } }},  \] which means that the second term in (\ref{extremeintegral}) can be neglected, leaving us with

\begin{equation}
1={a \over 2\pi N(1+af) } \left( {1 \over \pi^2+a^2s^2} \right) \ln \left( {\Lambda \over \epsilon} \right),
\end{equation}

\noindent its solution being

\begin{equation}
\Sigma(T)= T \sqrt{ {\alpha h \over 2\pi N (T+\alpha f)} - \pi^2}
\label{extremeintegralQED}
\end{equation}

\noindent where $h=\ln \frac{\Lambda }{ \epsilon} $.

Later we shall comment on this solution but first we turn our
attention to the mass cutoff case. Here the dominant contribution gives the equation 

\begin{equation}
1={a \over 2\pi N} \int_0^{\Lambda} {x dx \over (x^2+ afx^2+a \delta \ln2 )(x^2 +\pi^2+a^2s^2) }
\end{equation}

\noindent where again we have imposed the ultraviolet cutoff $\Lambda$. By neglecting all terms which are not divergent, when $\delta \rightarrow 0$, we get

\begin{eqnarray}
1={a \over 4 \pi N [(\pi^2+a^2s^2)(1+af)-a\delta \ln 2]} \ln \left[ {a  \delta \ln 2 +\Lambda^2(1+af) \over a \delta \ln 2} \right]
\end{eqnarray}

\noindent Now, if $\delta \ll \Lambda^2 f^2$ and $\delta \ll \frac{f \pi^2}{\ln{2}} $, which will certainly be the case for extremely small $\delta$, we
obtain the solution:

\begin{eqnarray}
\Sigma(T) &=& T \sqrt{ \frac{\alpha }{4 \pi N (T+ \alpha f )  } \ln \left[ \frac{\Lambda^2 (T +\alpha f)}{ \alpha \delta \ln2} \right] - \pi^2 } \nonumber \\
 &=& T \sqrt{ \frac{\alpha }{2 \pi N (T + \alpha f )}
 \left[ \ln \frac{\Lambda }{ \sqrt{ \delta \ln 2 }} + \ln \sqrt{ \frac{T}{\alpha } + f } \right] - \pi ^2}.
\label{extrememass}
\end{eqnarray}

\noindent Now we compare the two logarithms appearing in the second line of (\ref{extrememass}). Dynamical mass generation ($\Sigma \neq 0$) occurs for $T<T_c$ whereas $\Lambda$ is an ultraviolet cutoff. Furthermore $\alpha \geq e^2$ whereas $\delta$ is an infrared cutoff. f is just a constant. Thus it is reasonable to neglect the second logarithm.

\begin{equation}
\Sigma(T) \simeq T \sqrt{ \frac{\alpha \tilde{h} }{ 2 \pi N (T+ \alpha f) } - \pi^2 }
\label{approxextrememass}
\end{equation}

\noindent where $\tilde{h}=\ln \frac{\Lambda }{ \sqrt{\delta \ln 2}}$.

We see that (\ref{approxextrememass}) is very similar to
(\ref{extremeintegralQED}), fig.\ref{fig.1bQED}b and
fig.\ref{fig.1aQED}a respectively. We can therefore conclude that, for very small cutoff and provided it vanishes at $T=0$, 
 the choice of cutoff type has little effect on the solution. In both cases the infrared divergence is present in $\Sigma(T)$ as $h \ \mathrm{or} \ \tilde{h} \rightarrow \infty$.

We note that (\ref{extremeintegralQED}), (\ref{approxextrememass}) define  critical temperatures $T_c = \frac{\alpha h}{2 \pi^3 N} - \alpha f$, $T_c = \frac{\alpha \tilde{h}}{2 \pi^3 N} - \alpha f$ respectively.
However $\Sigma$ vanishes at $T=0$ as well. This is due to our having retained only the zeroth transverse mode: a finite value of $\Sigma$ at $T=0$ would be obtained, if we included the modes which are not infrared divergent. The real
 pathology of (\ref{extremeintegralQED}) or (\ref{approxextrememass})
 (apart from the infinity in the limit as $\epsilon$ or $\delta \rightarrow 0$) is that these solutions for $\Sigma$ increase as $T$ moves away from zero.
 This seems intuitively unnatural; the mass is analogous to an order parameter and we do not expect this to increase in magnitude as the temperature rises. In the following section we shall give a thermodynamic argument to support the idea that $\epsilon$ and $\delta$ should in fact be chosen to ensure that ${d \Sigma \over dT} \leq 0$ as $T \rightarrow 0$.
 We shall see that this can indeed be done, provided that they are
greater than some critical value.

Before proceeding, we add one further comment. The attentive reader might object to our previous remark about the mass being analogous to an order parameter, on the grounds that the Coleman-Mermin-Wagner theorem \cite{Coleman,Mermin} applies at finite temperature for a system in two spatial dimensions. This theorem forbids the existence of an order parameter which breaks a continuous symmetry of the system. However, the Kosterlitz-Thouless mechanism allows the global $\tau_3$-symmetry of QED$_3$ to remain unbroken despite the dynamical generation of mass. According to Witten \cite{WittenChiral}, the existence of dynamically generated fermion mass, which is the usual physical consequence of a non-zero order parameter, shows that even though $\langle \bar{\psi} \psi \rangle $  is not quite an order parameter, it is very close to being one. 

\section{Numerical Results}
\label{Numerical Results}

In this section we shall solve numerically equation (\ref{integralQED}) at $N=1$ for various values of $\delta$ and $\epsilon$.
 We shall see that for the specific values $\delta=\delta_c$ and $\epsilon=\epsilon_c$,
 $d \Sigma \over dT$ becomes negative for small $T/\alpha$. As we have
stated in the previous section, we believe our results to be
unphysical below $\delta_c$ and $\epsilon_c$; in the next section we
shall support this view by a simple thermodynamic argument. The behaviour of $ \Sigma $ as a function of $T$ is insensitive to
small variations of $N$ \cite{AitchisonEffect,DominicEffect}. The larger the $N$, the smaller the $ T_{c} $  and the more difficult it is to see the numerical solutions.  

In our numerical evaluation of (\ref{integralQED}), we must consider the
$T=0$ case as well as that of $T \neq 0$. To solve for $s(T=0)$ we use an equation which was derived in \cite{DominicEffect} for the zero temperature case of (\ref{rearranged}), 
retaining only the longitudinal part:

\begin{equation}
1={Q(\infty,s) \over 2 \pi N} \equiv {2 \over (2 \pi)^2 N} \left[
\left(1-{(0.125)^2 \over (0.125)^2+s^2} \right){\pi \over 2s}-{0.125
\over (0.125)^2 +s^2} \ln \left( {s \over 0.125} \right) \right].
\label{0integral}
\end{equation}

\noindent Before using this equation, we need to replace $N$ by $N/2$. This is because (\ref{0integral}) was only derived for the longitudinal part of (\ref{SD}). Since at $T=0$ the transverse and the longitudinal contributions are
identical we need only double the longitudinal contribution.

Evaluating (\ref{0integral}) numerically yields a value of
$s(T=0)={\Sigma(T=0) \over \alpha}=0.0578$; note that this result as
well as (\ref{0integral}) itself are cutoff independent, due to the fact that both cutoffs go to zero at $T=0$, where there is no infrared divergence. There is one more crucial property of (\ref{0integral}); $\lim_{T \rightarrow
0}  Q_0(a) =\infty$, where $Q_0(a)=Q(a,s=0)$. Combining this result
with some other considerations in Appendix B, will lead to 
conclusions about the critical temperature and about some properties of a solution $s=s(T)$ of the S-D equation.

To solve (\ref{integralQED}) numerically at $T \neq 0$ for both $\delta$ and $\epsilon$ dependence, we choose a value of $\delta$ or $\epsilon$ and 
seek to determine $1/a_c \equiv {T_ck_B \over \alpha}$. 
To do this we know from Appendix B that $Q_0(a_c)=2 \pi$ at
$N=1$. Therefore we start by varying $Q_0$ as a function of $a$ until
the condition $Q_0=2 \pi$ is satisfied, the value of $a$ at that point
being $a_c$. Once we have determined $a_c$, we start looking at values
of $a>a_c$ for which $Q_0>2 \pi$. For $Q_0< 2\pi$ we know from Appendix B that $\Sigma=0$.
  At each value of $a$ we consider, we vary $Q(a,s)$ as a function of $s$ until (\ref{integralQED}) is satisfied, namely such that $Q(a,s)=2 \pi$ for N=1.
  Using each value of $s=\Sigma/\alpha$ we evaluate for a given $a$,  
as well as for the $T=0$ solution, we can construct full temperature
dependent solutions of (\ref{integralQED}) for a particular value of $\epsilon$ or $ \delta $.

The first set of numerical solutions are those for the integral cutoff
method, as shown in fig.\ref{fig.2QED}. For each of these solutions we give a value of ${k_BT_c \over \alpha}$ and we calculate the ratio $r ={2 \Sigma(T=0) \over T_ck_B}$.
 As one can see for all values of $\epsilon$, except for
$\epsilon=2.5$ and $\epsilon=2$, the gradient ${d \Sigma \over dT}$ is positive for small $T/\alpha$. As $\epsilon$ decreases we see that this effect becomes more pronounced.
 We also see that as $\epsilon$ decreases, $r$ decreases (as shown in table \ref{table1}); this is due to the increase in the critical temperature, while at the 
same time still retaining a cutoff independent $T \rightarrow 0$
limit. Both these effect are consistent with the solutions of the
 previous section for extremely small $\epsilon $, where we have found
that $T_c \simeq {h \alpha \over 2 \pi N}$, with $h= \ln \left( {
\Lambda \over \epsilon } \right) $. Also we see for these
extremely small $\epsilon$ solutions, that ${d \Sigma \over dT}
\rightarrow \infty$ as $T \rightarrow 0$ (see fig.\ref{fig.1aQED}a).

The second set of numerical solutions are those using a mass cutoff and are shown in fig.\ref{fig.3QED}.
 We give values of $r$ and $k_BT_c$ for each solution of a specific $\delta$ in table \ref{table2}.
 It is important to notice that again there is a transition: below a certain
 value of $\delta$ we see that ${d \Sigma \over dT}$ is positive for
small $T/\alpha$. A positive gradient case is shown for $\delta=0.05$. Again we notice that the $r$-values fall with decreasing $\delta$,
 consistent with the extremely small $\delta$ solution  of the previous section. One should notice that for the mass cutoff,
 if we look only at $r$ values for solutions
 which do not have a positive gradient, we see that $r$ changes slowly, namely by a factor of order $0.7$
 over an order of magnitude in $\delta$ ($\delta$ ranging from 2 to
0.25). We see that $r=6.2$ for the value of $\delta=0.5$, which is very close to the value of $r$ given in \cite{DominicEffect} of $6.17$.
This is because $\delta$ is at a value close to the corresponding longitudinal plasmon mass of ${2 \ln 2 \over \pi}a$, 
so making the transverse contribution roughly the same as the longitudinal, as in \cite{AitchisonBeyond}.

\begin{table}[p]
\begin{center}
\begin{tabular}{|@{\hspace{1.25em}}c@{\hspace{1.25em}}*{6}{|c}|} \hline
& & & & & & \\
$\epsilon$ & 2.5 & 2 & 1.5 & 1 & 0.5 & 0.25 \\
& & & & & & \\ \hline
& & & & & & \\
$T_c$ & 0.016 & 0.018 & 0.023 & 0.026 & 0.035 & 0.0445 \\
& & & & & & \\ \hline
& & & & & & \\
$r$ & 7.23 & 6.42 & 5.03 & 4.45 & 3.30 & 2.60 \\
& & & & & & \\ \hline
\end{tabular}
\caption{The values of \( r = \frac{2 \Sigma (T=0) }{k_B T_c} \) for
different values of the integral cutoff $ \epsilon $. $ \Sigma $ is the dynamically generated mass and $ T_c $ the critical temperature measured in units \( \frac{e^2}{k_B} \). For the allowed region of values
 $ \epsilon \geq \epsilon_c $, $ \epsilon_c \simeq 2 $, the average value of  $r$ is about 6.8 in agreement with previous papers. \label{table1}}
\end{center}
\end{table}

\begin{table}[p]
\begin{center}
\begin{tabular}{|@{\hspace{1.25em}}c@{\hspace{1.25em}}*{5}{|c}|} \hline
& & & & & \\
$\delta$ & 2 & 1 & 0.5 & 0.25 & 0.05 \\
& & & & & \\ \hline
& & & & & \\
$T_c$ & 0.015 & 0.0164 & 0.0185 & 0.0215 & 0.031 \\
& & & & & \\ \hline
& & & & & \\
$r$ & 7.70 & 7.05 & 6.25 & 5.38 & 3.73 \\
& & & & & \\ \hline
\end{tabular}
\caption{The values of \( r = \frac{2 \Sigma (T=0) }{k_B T_c} \) for
different values of the mass cutoff $ \delta $. $ \Sigma $ is the dynamically generated mass and $ T_c $ the critical temperature measured in units \( \frac{e^2}{k_B} \). For the allowed region of values
 $ \delta \geq \delta_c $, $ \delta_c \simeq 0.2 $, the average value of $r$ is about 6.6 in agreement with previous papers. \label{table2}}
\end{center}
\end{table}

\section{A thermodynamic argument for ruling out $\delta<\delta_c$ and $\epsilon<\epsilon_c$ solutions}

We now look for a simple thermodynamic argument to rule out solutions
with positive gradient at small $T/\alpha$. We consider a gas of fermions, which we take to be dilute, so that we may ignore any interaction term.
 The chemical potential of a dilute gas is negligible and may be
discarded. We shall also assume the mass to be temperature
dependent. The partition function takes the usual form for a
non-interacting gas of fermions (with $\mu =0$), except that now 
$\omega=({\bf p}^2+\Sigma^2)^{1/2}$ is temperature dependent:

\begin{equation}
\ln Z = 2VN \int {d^2 {\bf p} \over (2 \pi)^2} \left[ \beta \omega (T)+ 2 \ln \left( 1 + e^{- \beta \omega (T)} \right) \right]
\label{partition}
\end{equation}

\noindent where $V$ is the volume of the system and $N$ is the number
of flavours. We shall calculate the entropy of this gas. It is convenient to divide $\ln Z$ into two contributions, $\ln Z=\ln Z_G+\ln Z_E$, where $\ln Z_G$ is the contribution from the ground state and  $\ln Z_E$ is the contribution from the excited states. We first consider the ground state 
contribution. We calculate the frequency integral imposing an ultraviolet
frequency cutoff, $\Lambda_G$, since the integral in $\ln Z_G$ is ultraviolet divergent.  We then have

\begin{equation}
\ln Z_G = {VN \beta \over  \pi} \left[ {\Lambda^3_G \over 3}-{\Sigma^3(T) \over 3} \right].
\label{groundpartition}
\end{equation}

\noindent $\Lambda^3_G$ should be thought as an (infinite) constant
arising from the zero-point energy contribution at $T=0$; since it is constant it will not affect the entropy. We know that the entropy is related to the partition function
 by $S={\partial T \ln Z \over \partial T}$ (in these calculations we
shall set $k_B=1$). From this we find that the entropy of the ground state
 takes the form:

\begin{equation}
S_G= -{VN \over  \pi} \Sigma^2(T) {d \Sigma \over dT}
\label{groundentropy}
\end{equation}

\noindent We now look at the contribution arising from excited fermion states. To perform the integral analytically we expand out the logarithm in (\ref{partition})
 for $\beta \Sigma(T) >1$ retaining only the first term. This is
allowed, because we are interested only for $T/\alpha$ small and
$\Sigma(T=0)/\alpha \neq 0$. Thus we arrive at the following form for the $\ln Z_E$ term:

\begin{equation}
\ln Z_E = {4VT^2N \over 2 \pi} \left( { \Sigma \over T} + 1 \right) 
\ e^{- \frac{ \Sigma (T)}{T}}.
\label{excitedpartition}
\end{equation}

\noindent It follows that the corresponding
 entropy, $S_E$ , at small $T/\alpha$ is given by

\begin{equation}
S_E=  {4VNT^2 \over 2 \pi} \left( {3 \Sigma(T) \over T}+3+{\Sigma^2(T) \over T^2}-{\Sigma \over T}{d \Sigma \over dT} \right) \  e^{- \frac{ \Sigma (T)}{T}}.
\label{excitedentropy}
\end{equation}

\noindent For very large $\Sigma/T$ we notice that
(\ref{excitedentropy}) is suppressed by a factor of $\exp (-\Sigma
/T)$, so at large $\Sigma(T)/T$ this contributes little to the entropy
of the system. It is $S_G$ which is the dominant factor. We therefore
conclude that, although our model may be an over-simplification of a real physical system in QED$_3$, positive 
${d \Sigma \over dT}$ at small $T/\alpha$ leads to negative entropy,
which is clearly unphysical. The consequence of this is that there is
a boundary between physical and non-physical solutions in the
parameter spaces of the two types of solution. From this we may
conclude that the contribution from higher orders in $1/N$, if indeed
it regulates the infrared divergence, must be above a certain magnitude.

There is one point that needs to be addressed before we finish this section. If ${d \Sigma \over dT} \neq 0$ at $T=0$, (\ref{groundentropy})
 implies that $S \neq 0$ at $T=0$, which is unphysical: when $T \rightarrow 0$ the 3rd law states that $S \rightarrow 0$. From our
 numerical data we strongly suspect that for all values of $\delta$ and $\epsilon$ (except $\delta=0$, $\epsilon=0$)
 ${d \Sigma \over dT} =0$ at T=0, when all modes are considered. Although we have been unable to prove this, we are able to present
 an argument showing that $  \left. {d \Sigma \over dT } \right|_{T=0}
$ is independent of $\epsilon$ and $\delta$ in  Appendix B, so that $
\left. { d \Sigma \over dT} \right|_{T=0}$ is the same in all cases
(except when $\epsilon(\delta)=0$ ). We show in fig.\ref{fig.4QED} what we think happens as $T \rightarrow 0$, where the gradient of all the curves goes to zero as $T \rightarrow 0$ in
 accordance with the 3rd law. Although this behaviour is likely to be true for very small $T/\alpha$, $d \Sigma \over dT$ is 
significantly
 positive at small enough $T/\alpha$ for $\epsilon<\epsilon_c$ and $\delta<\delta_c$, so that $\epsilon_c$ and $\delta_c$ are good 
estimates of the boundary between physical and non-physical regions;
the contribution from (\ref{excitedentropy}) is very small, when compared
 to (\ref{groundentropy}) in the region where we see positive $d \Sigma \over dT$.

\section{An approximate form for $\Sigma$ and estimates of $\delta_c$ and
$\epsilon_c$}

In this section we obtain an analytic representation of $\Sigma$
as a function of $T$ and $\epsilon (\delta)$. This will give us a simple way 
of estimating $\epsilon_c$ ($\delta_c$) and it may also be useful in 
applications to phenomenology, for example to a calculation of the 
temperature dependence of the Meissner effect
 (although we shall see later that a mass cutoff is not consistent
with the $U(1) \times U(1)$ model of superconductivity,
\cite{DoreyQED}). From our numerical solutions (fig.\ref{fig.2QED}
and fig.\ref{fig.3QED} ) we
see that $\Sigma$ as a function of $T$ looks very much like the
section of an ellipse, centered about some value $T_0$, in the
quadrant $\Sigma>0$ and $T>0$. Such an ellipse takes the form (see
fig.\ref{fig.5aQED}a, fig.\ref{fig.5bQED}b)

\begin{equation}
m^2(T)=1+{2T_0T \over T'^2_c-2T_0T'_c}-{T^2 \over T'^2_c-2T_0T'_c}
\label{ellipse}
\end{equation}

\noindent in terms of the two parameters $T_0$ and $T'_c$ and
where $m(T)= {\Sigma(T) \over \Sigma(T=0)}$.
 Fitting (\ref{ellipse}) to the numerical data, we have found that it is better to use both $T_0$ and $T'_c$ to fit $\Sigma(T)$, 
instead of fixing the value of
 $T'_c$ at the calculated  $T_c$ for a particular $\epsilon$
($\delta$) and varying only $T_0$. To indicate the accuracy of the fit
we show in fig.\ref{fig.6aQED}a and fig.\ref{fig.6bQED}b both the
approximation and the numerical data for one chosen value of $\epsilon$ and one of $\delta$. Fitting both $T_0$ and $T'_c$ we are able to build up graphs of $T_0$ for both cutoffs (see fig.\ref{fig.7aQED}a and fig.\ref{fig.7bQED}b).
Using the simple condition that $T_0 \leq 0$ implies $ {d \Sigma \over dT }\leq 0$, we are able to estimate $\epsilon_c$ and $\delta_c$; we find $\epsilon_c \sim 2$ and $\delta_c \sim 0.2$.

\section{Discussion}
 
Although in QED$_3$ both types of cutoff lead to roughly the same behaviour for the mass gap, as a function of $T$,
 the mass cutoff is ruled out, if applied to the $U(1) \times U(1)$ model of high-$T_c$ superconductivity \cite{DoreyQED}. The problem is that the transverse
 part of $\Delta_{\mu \nu}$ (the statistical gauge field propagator) in this model must be massless. If $\Delta_{\mu \nu}$ has
 a transverse mass, it no longer generates (via an analogue of the Higgs mechanism)
 a transverse mass in $D_{\mu \nu}$, the electromagnetic propagator and there is therefore no Meissner effect. This also suggests 
that infrared regularizing contributions from beyond leading order in $1/N$
 may come from the vertex function in (\ref{SD}), if QED$_3$ is indeed a model of high-$T_c$ superconductivity.

Although we found (Appendix A)
that the transverse contribution to the photon propagator still
behaves at $p_{0b}=0$ for small $|{\bf p}|$ as $1/{\bf p}^2$, 
even if massive fermions are used in the vacuum polarization,
 our expressions for $\Pi^{\Sigma}_{\mu \nu}$ remain useful. It will be interesting to see what effect a photon propagator 
involving $\Pi^{\Sigma}_{\mu \nu}$ (instead of $\Pi_{\mu \nu}$) has on the calculation of a $3$-momentum independent mass, $\Sigma$.
 By looking at (A.21), we see that as $\beta \Sigma \rightarrow \infty$, the plasmon mass tends to zero in the longitudinal part of 
the photon propagator at $p_{0b}=0$. It may be that instead of monotonically decreasing with $\beta \Sigma$, at large enough
 $\beta \Sigma$, the contribution from the zeroth longitudinal mode to
the gap equation is enhanced. If this behaviour 
is present in the zeroth longitudinal mode, it might cause
(\ref{integralQED}) to favour smaller values of $\Sigma$ than would be
expected, if, as in the case of $\Pi_{\mu \nu}$, the zeroth mode plasmon mass contribution was independent of $\Sigma$.

It might also be interesting in the context of the infrared problem to look at the effects of the wavefunction renormalization
 given in \cite{AitchisonEffect}, where $ \mathcal{M} $, the physical mass, is related to $\Sigma$ by
\begin{equation}
{\cal M}(p_f)={\Sigma(p_f) \over 1+A(p_f)}.
\end{equation}
We note that  the integrand in the equation for $A(p_f)$ has a pole at $q_{b}=p_f-k_f=0$, where $k_f$ is the loop momentum
 (see equation (6) of \cite{AitchisonEffect}). We strongly suspect that this will lead effectively to an infrared divergence, which will then dominate 
the equation for $A(p_f)$ for very small cutoff. For the case of a 3-momentum independent $ \mathcal{M}$, one should be able to
 derive an expression relating $A(p_f)$ to $ \mathcal{M}$. From this one could see what effect $A(p_f)$ has on $ \mathcal{M} $ for
 small cutoffs, the hope being that $ \mathcal{M}$ may be finite for $T \neq 0$ without a cutoff. One should note, however, that 
we may have to enforce the Ward-identities at finite temperature, which the wavefunction renormalization approach in \cite{AitchisonEffect} does not do, in order to regulate the infrared divergence.

\begin{center}
  {\bf  ACKNOWLEDGMENTS }
\end{center}
We wish to thank our supervisor I.J.R. Aitchison for his help and we
acknowledge financial support from PPARC. We also thank  Adrian
Campbell-Smith for help in proof reading Appendix-A. 

\newpage

\renewcommand{\theequation}{\mbox{A.\arabic{equation}}}
\renewcommand{\thefigure}{\mbox{A.\arabic{figure}}}

\setcounter{equation}{0}

\begin{center}
{\bf Appendix A - Calculation of the polarization tensor with a fermion mass to leading order in $1/N$.}
\end{center}

In \cite{DominicEffect} it was found, for the polarization tensor
$\Pi_{\mu \nu}$  calculated for massless fermions, that there was an
infrared divergence in the Schwinger-Dyson equation for the fermion
mass. This infrared divergence comes from the zeroth frequency part of
the photon propagator which goes like $1/ {\bf p}^2$ for small momenta at $T \neq 0$. In this appendix we shall calculate the polarization tensor $\Pi^{\Sigma}_{\mu \nu}$ with fermion propagators of
 mass $\Sigma$; we use the superscript $\Sigma$ to distinguish this
polarization tensor from that in the massless case. Again this
calculation is to be done at finite $T$ and in the Matsubara
formalism. It is precisely the low-$|{\bf p}|$ behaviour of the resultant photon propagator, $\Delta^{\Sigma}_{\mu \nu}$ at $p_{0b}=0$, that we shall be interested in; we  want to see if the infrared divergence is regulated by the replacement of the massless fermion propagators with massive  fermion propagators in our calculation of $\Pi^{\Sigma}_{\mu \nu}$.

To calculate  $\Pi^{\Sigma}_{\mu \nu}$ to leading order in $1/N$  we
need only consider the contribution from the diagram shown in fig.A1.

\begin{figure}
\begin{center}
\resizebox{16.5cm}{!}{\includegraphics*[175pt,365pt][423pt,505pt]{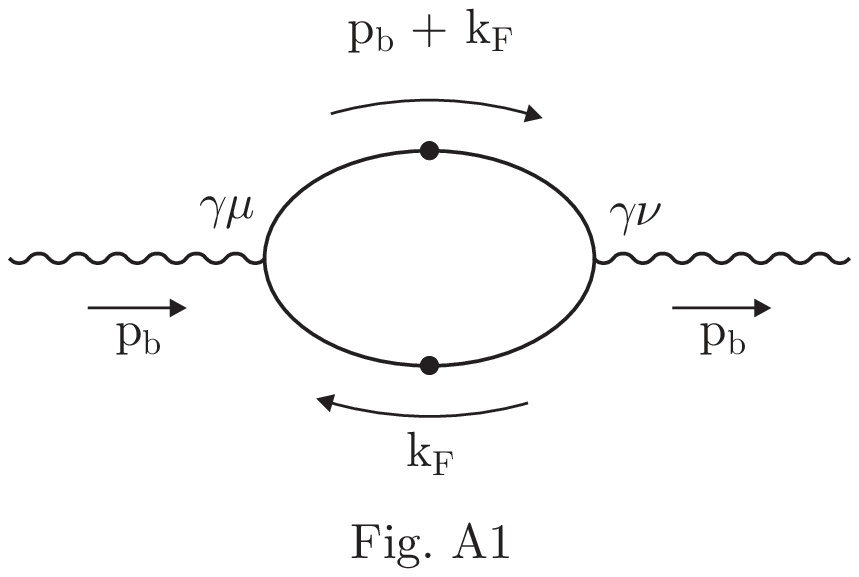}}
\caption{The vacuum polarization tensor $\Pi^{\Sigma}_{\mu \nu}$.}
\end{center}
\end{figure}

From fig.A1 we deduce that the polarization tensor takes the following form
 
\begin{equation}
 \Pi^{\Sigma}_{\mu \nu}(p_b) = {\alpha \over \beta} \sum_n \int {d^2k \over (2\pi)^2}
 \mathrm{tr} \left( \gamma_{\mu} {1 \over p_b \!\!\!\!\! / +  k_f \!\!\!\!\! / + \Sigma(k_f+p_b)}
\gamma_{\nu} {1 \over k_f \!\!\!\!\! / + \Sigma(k_f)} \right).
\end{equation}

In our calculation of $\Pi^{\Sigma}_{\mu \nu}$ we take $\Sigma$ to be
constant with respect to 3-momentum. If $\Sigma$ was to vary with
respect to frequency or momentum, we would at this point need to insert a variational ansatz and solve
integral equations for the variational parameters derived from the Schwinger-Dyson equation for the mass, $\Sigma$; or we would have to treat (A.1) as a separate integral equation. By use of Feynman parameterization and other manipulations we arrive at the following expression
 for  $\Pi^{\Sigma}_{\mu \nu}$:
\begin{eqnarray}
\Pi^{\Sigma}_{\mu \nu} = {4 \alpha \over \beta} \int^1_0 dx \sum_n \int {d^2l \over (2\pi)^2} {2l_{\mu}l_{\nu} \over (l^2+p^2_bx(1-x)+\Sigma^2)^2} 
-{ \delta_{\mu \nu} \over (l^2+p^2_bx(1-x)+\Sigma^2)} \nonumber \\
+{2(x(1-x))(p_b^2 \delta_{\mu \nu} - p_{\mu b}p_{\nu b}) \over (l^2+p^2_bx(1-x)+\Sigma^2)^2} 
+{(1-2x)(l_{\mu}p_{\nu b}+l_{\nu}p_{\mu b}-(p_b.l)\delta_{\mu \nu}) \over (l^2+p^2_bx(1-x)+\Sigma^2)^2 } 
\end{eqnarray}
where $l$ is the shifted 3-momentum, related to $k_f$ by
$l=k_f+xp_b$. For our purposes we need only consider $\Pi^{\Sigma}_{ij}$ and $\Pi^{\Sigma}_{00}$.

 Before proceeding any further there are some immediate points that need to be raised. The first is that, if we were to take the limit $T \rightarrow 0$ of (A.2), the first two terms would cancel after regularization. This is because the first two terms have the same tensor structure, namely both are proportional to $\delta_{\mu \nu}$, and using
 dimensional regularization one is able to show that these two terms are equal in magnitude and opposite in sign. But now consider the case when $T \neq 0$: in this case we have a sum over $k_{0f}$ components in $l_0$ instead of an $l_0$-integral and so 
there is a preferred direction in $3$-momentum space. The upshot of this is that the dimension of the $l$-integrations is effectively reduced  and the first term is no longer proportional to $\delta_{\mu \nu}$.

In the case of $\Pi^{\Sigma}_{ij}$  the first two terms
in (A.2) do cancel at $T \neq 0$, due to the $l$-integrals
 having the same form as those for $T=0$, although now two dimensional instead of three dimensional.
For the $\Pi^{\Sigma}_{00}$ terms the situation is slightly trickier, for the first term of (A.2) instead of having $l_il_j$ as a 
factor has $l_0^2$. It is this term in $\Pi^{\Sigma}_{00}$ which causes the first term to be no longer proportional to $\delta
_{\mu \nu}$, and this means that for $\Pi^{\Sigma}_{00}$
the first two terms in (A.2) no longer cancel.

At this point we should ask ourselves the question: is there any constant, leading order term when we set $p_b=0$? If we look 
at  $\Pi^{\Sigma}_{ij}$ , the answer is no, for when we set $p_b=0$ only the first two terms survive, which cancel. For $\Pi^{\Sigma}_{00}$
, the answer is yes, for although only the first two terms survive, they do not cancel. This is precisely why, as
 we shall see in our calculation of $\Pi^{\Sigma}_{\mu \nu}$, there is
 a plasmon mass only in the longitudinal part of the photon propagator
at $p_{0b}=0$, ${\bf p}=0$. Exactly the same argument holds for
$\Pi_{\mu \nu}$, the massless case. Although this is true, we are
still interested to see what is the actual low-$|{\bf p}|$ behaviour
the $p_{0b}=0$ part of the photon propagator.

Continuing our analysis, we write both $\Pi^{\Sigma}_{00}$ and $\Pi^{\Sigma}_{ij}$ as:
\begin{equation}
\Pi^{\Sigma}_{00} = {4\alpha \over \beta} \int^1_0 dx \int {d^2l \over (2\pi)^2}
[ S^{\Sigma}_1 +2({\bf l}^2+\Sigma^2+x(1-x)p^2_{0b})S^{\Sigma}_2+
(1-2x)p_{0b}S^{* \Sigma}_2]
\end{equation}
and
\begin{equation}
\Pi^{\Sigma}_{ij} = {4\alpha \over \beta} \int^1_0 dx \int {d^2l \over (2\pi)^2}
[ 2x(1-x)(p^2_{b}\delta_{ij}-p_ip_j)S^{\Sigma}_2+
(1-2x)p_{0b}S^{* \Sigma}_2 -\delta_{ij}(1-2x)p_{0b}S^{* \Sigma}_2]
\end{equation}
where
\begin{eqnarray}
S^{\Sigma}_i = \sum^{\infty}_{n=-\infty} {1 \over (l^2_0+{\bf l}^2+\Sigma^2+x(1-x)p^2)^i } \nonumber \\
S^{* \Sigma}_i = \sum^{\infty}_{n=-\infty} {l_0 \over (l^2_0+{\bf l}^2+\Sigma^2+x(1-x)p^2)^i }
\end{eqnarray}
 From (A.5) one can easily deduce the following relationships
\begin{eqnarray}
&&S^{\Sigma}_2=-{1 \over 2y}{\partial S^{\Sigma}_1 \over \partial y}=-
{1 \over 2|{\bf l}|}{\partial S^{\Sigma}_1 \over \partial |{\bf l}|}=
-{1 \over 2x(1-x)p_b}{\partial S^{\Sigma}_1 \over \partial p_b} \\
&&S^{*\Sigma}_2 = -{\beta \over 2}{\partial S^{\Sigma}_1 \over \partial \omega} \end{eqnarray}
where $y^2={\bf l}^2+\Sigma^2+x(1-x)p^2$ and $\omega =\pi(2mx+1)$. Using (A.6) and (A.7) we are able to 
express $\Pi^{\Sigma}_{00}$ and $\Pi^{\Sigma}_{ij}$:
\begin{equation}
\Pi^{\Sigma}_{00}={4\alpha \over \beta} \int^1_0 dx \int {d|{\bf l}| \over (2\pi)} \left[|{\bf l}| S^{\Sigma}_1+(|{\bf l}|^2+\Sigma^2) {\partial S^{\Sigma}_1  \over \partial |{\bf l}| }+{p^2_0 \over p}{\partial S^{\Sigma}_1  \over \partial p }|{\bf l}|-(1
-2x){\beta p_{0b} \over2}{\partial S_1 \over \partial \omega} \right]
\end{equation}
\begin{equation}
\Pi^{\Sigma}_{ij}={4\alpha \over \beta} \int^1_0 dx \int {d|{\bf l}| \over (2\pi)} \left[ {-(p^2_b \delta_{ij}-p_ip_j) \over p}{\partial S^{\Sigma}_1  \over \partial p }|{\bf l}|+(1-2x){\beta p_{0b} \over2}{\partial S_1 \over \partial \omega}\delta{ij} \right]
\end{equation}
We can evaluate the sum $S^{\Sigma}_1$  by the usual complex integration, so obtaining the result
\begin{equation}
S^{\Sigma}_1 = {\beta \over 4iy}[\cot (\frac{\omega+i\beta y}{2})+ \cot (\frac{\omega-i\beta y}{2})]
\end{equation}
Using (A.8), (A.9) and (A.10), as well as regularizing the integrals so
that the ultraviolet singularities cancel, one is able to evaluate the
$|{\bf l}|$-integral. This gives us final expressions for
$\Pi^{\Sigma}_{00}$ and $\Pi^{\Sigma}_{ij}$,
\begin{eqnarray}
&&\Pi^{\Sigma}_{00} = \Pi^{\Sigma}_3 -{p^2_0 \over p_b^2}\Pi^{\Sigma}_1 -\Pi^{\Sigma}_2+\Pi^{\Sigma}_4 \\
&&\Pi^{\Sigma}_{ij} = \Pi^{\Sigma}_1(\delta_{ij}-\frac{p_ip_j}{p^2_b})+\Pi^{\Sigma}_2 \delta_{ij} 
\end{eqnarray}
where
\begin{eqnarray}
&&\Pi^{\Sigma}_1 = {\alpha p^2_b \over 2\pi} \int^1_0 dx {x(1-x) \over (x(1-x)p^2_b + \Sigma^2)^{1/2} }{ \sinh( \beta(p^2_bx(1-x)+\Sigma^2)^{1/2}) \over D^{\Sigma}_m(x,p_b,\beta) } \nonumber \\
&&\Pi^{\Sigma}_2 = \left( {\alpha m \over 2 \beta} \right) \int^1_0 (1-2x)dx {\sin(2xm\pi) \over D^{\Sigma}_m(x,p_b,\beta) } \\
&&\Pi^{\Sigma}_3 ={\alpha \over \pi \beta} \int^1_0 dx \ln(4D^{\Sigma}_m (x,p_b,\beta)) \nonumber \\
&&\Pi^{\Sigma}_4=-{\alpha \beta \Sigma^2 \over 2\pi} \int^1_0 {dx \sinh(\beta(p^2_bx(1-x)+\Sigma^2)^{1/2}) \over (p_b^2(1-x)+\Sigma^2)^{1/2}
D^{\Sigma}_m(x,p_b,\beta) } \nonumber
\end{eqnarray}
and
\begin{equation}
D^{\Sigma}_m(x,p_b,\beta)= \cosh^2 \left( {\beta(p_b^2x(1-x)+\Sigma^2)^{1/2} \over 2} \right) -\sin^2 \pi mx
\end{equation}

 We now examine whether the expressions (A.11), (A.12), (A.13) and (A.14) give the correct limits. We can easily
 show that $\lim_{\Sigma \rightarrow 0} \Pi^{\Sigma}_{ij} =\Pi_{ij}$ and  $\lim_{\Sigma \rightarrow 0} \Pi^{\Sigma}_{00} =\Pi_{00}$,
 where $\Pi
_{ij}$ and $\Pi_{00}$, the results for the massless calculation, are
given in \cite{DoreyQED,DominicEffect}. When $\Sigma \rightarrow
\infty$ we require that $\Pi^{\Sigma}_{ij} \rightarrow 0$ and $\Pi^{\Sigma}_{00}
\rightarrow 0$; by looking at (A.1), we can see that this requirement 
 must be true. Also to be consistent with the results at zero temperature \cite{AitchisonBeyond} these terms must vanish in the $\Sigma \rightarrow \infty$ limit (although this does not rule out the presence of a term independent of $\Sigma$ and proportional to 
 $\alpha / \beta$ which vanishes in this limit). For $\beta\Sigma \gg  \beta p_b$ and $\beta\Sigma \gg  1$ we have the following:
\begin{eqnarray}
\Pi_1 \simeq {\alpha p^2_b \over 12 \Sigma \pi} \; &\Pi_2& \simeq {2\alpha \over \beta \pi}e^{-\sigma \beta}, p_{0b} \neq 0 \\
\Pi_3={\beta \Sigma \over \pi} \left( {\alpha \over \beta} \right) \; &\Pi_4& =-{\beta \Sigma \over \pi} \left( {\alpha \over \beta} \right) .\nonumber
\end{eqnarray}
 From (A.15) it is easy to show that
$\Pi^{\Sigma}_{00}$ and  $\Pi^{\Sigma}_{ij}$ have the correct $\Sigma \rightarrow  \infty$ limit.
 
We now need to deduce the low $|{\bf p}|$-behaviour for the photon propagator. In the Landau gauge we know that the photon
 propagator must take the form: 
\begin{equation}
\Delta^{\Sigma}_{\mu \nu} ={A_{\mu \nu} \over p_b^2+\Pi^{\Sigma}_{A}(p_b)}+
{B_{\mu \nu} \over p_b^2+\Pi^{\Sigma}_{B}(p_b)}
\end{equation}
where $A_{\mu \nu}$ and $B_{\mu \nu}$, the longitudinal and transverse
projection operators, are given, to leading order in $1/N$, by
\begin{eqnarray}
&&A_{\mu \nu}=\left( \delta_{\mu0}-{p_{\mu}p_0 \over p^2} \right) \left( {p^2 \over {\bf p}^2 } \right) \left( \delta_{0\nu}-{p_0p_{\nu} \over p^2} \right) \\
&&B_{\mu \nu}=\delta_{\mu i} \left( \delta_{ij}-{p_ip_j \over 
{\bf p}^2 }\right) \delta_{j \nu}.
\end{eqnarray}

 $\Pi_A^{\Sigma}$ and  $\Pi_B^{\Sigma}$ are 
related to $\Pi^{\Sigma}_{ij}$ and $\Pi^{\Sigma}_{00}$ by the following formulae (for further discussion see \cite{DoreyQED,DominicEffect}):

\begin{equation}
\Pi^{\Sigma}_A = \Pi^{\Sigma}_{00} {p_b^2 \over {\bf p}^2} \;\;\; 
\Pi^{\Sigma}_B =\Pi^{\Sigma}_{ii} -\Pi^{\Sigma}_{00} {p_{0b}^2 \over {\bf p}^2}  \end{equation}
Using (A.19) we can deduce the zero frequency forms of
$\Pi^{\Sigma}_A$ and $\Pi^{\Sigma}_B$ in terms of $\Pi^{\Sigma}_1$,
$\Pi^{\Sigma}_3$ and $\Pi^{\Sigma}_4$. Noting that at $p_{0b}$, $\Pi^{\Sigma}_2=0$ for all $|{\bf p}|$,
\begin{equation}
\Pi^{\Sigma}_A(p_{0b}=0) =\Pi^{\Sigma}_3 +\Pi^{\Sigma}_4 \;\;\; 
\Pi^{\Sigma}_A(p_{0b}=0) =\Pi^{\Sigma}_1 
\end{equation} 

From (A.11), (A.12), (A.13) and (A.20) we find that to leading order in 
$|{\bf p}| $:
\begin{equation}
\Delta^{\Sigma}_{\mu \nu}={A_{\mu \nu} \pi \beta \over 2 \alpha \ln(2\cosh(\beta \Sigma/2))-\beta\Sigma \alpha \tanh(\beta\Sigma)}+
{B_{\mu \nu} 12 \pi \Sigma \over 2 {\bf p}^2(12\pi \Sigma+\alpha \tanh(\beta\Sigma))}
\end{equation}
Note that unlike (A.15), we have not taken $\beta \Sigma \gg 1$, but rather 
 $\beta \Sigma \gg \beta |{\bf p}|$ at $p_{0b}=0$; this gives us factors of $\tanh \beta \Sigma$ and $\ln (2 \cosh \beta \Sigma /2)$.
 From (A.21) one can plainly see that
$\Delta^{\Sigma}_{\mu \nu}$ goes as $1/{\bf p}^2$ in the transverse part for
 small $|{\bf p}|$. This is exactly the same behaviour as $\Delta_{\mu \nu}$, calculated with massless fermion propagators; therefore the inclusion of the finite fermion mass in the vacuum polarization does not remove the infrared divergence in the gap equation.

\renewcommand{\theequation}{\mbox{B.\arabic{equation}}}

\setcounter{equation}{0}

\begin{center}
{\bf Appendix B - Analytic properties of the solution to the gap equation.}
\end{center}

In this appendix we are interested in those properties of the 
 solution $\Sigma=\Sigma(T)$ of the S-D equation in the form (\ref{integralQED})  which can be found analytically. In the analysis that follows, calculations will be given for the integral cutoff case, but the properties found will be true
of the mass cutoff case also. 
For any finite number of flavours $N$, we shall prove the following points:

(a) There is a non-zero critical temperature $T_c$.

(b) $T_c$ is unique.

(c) For $T>T_c$, $s$ is identically zero.
 
(d) At $T=T_c$, $\frac{ds}{dT}  \rightarrow -\infty $.

(e) At $T=0$, $\frac{ds}{dT}$ is cutoff independent.

\noindent It should be mentioned in passing that all these proofs apply also to the solution for constant $\Sigma$, where only the longitudinal part of (\ref{integralQED}) is retained as in \cite{DominicEffect}.

We define $Q_{0}(T) \equiv Q(T,s^2=0)$. For the purposes of this appendix we first need to prove that $\frac{dQ_{0}(T)}{dT} < 0$ and $\frac{\partial Q(T,s^2)}{\partial s^2} < 0$. 

To show $\frac{dQ_0}{dT} < 0$, instead of using equation (\ref{integralQED}) it is much easier to go back to (\ref{rearranged}), (\ref{transverse}) and define $Q_0(T)$ as:

\begin{eqnarray}
Q_0(T) \equiv a \sum^{\infty}_{m=1} \int^{\infty}_{\epsilon}{2x dx \over x^2+(2\pi m)^2+0.125a [x^2+(2\pi m)^2]^{1/2} }\nonumber \\
\times \left[ {1 \over x^2+[2\pi(m+1/2)]^2}+{1 \over x^2+[2\pi(m-1/2)]^2} \right] \nonumber \\
+\int^{\infty}_{\epsilon} {ax dx \over x^2+\pi^2} \left[ {1 \over x^2+\beta^2\Pi^1_0 }+ {1 \over x^2+\beta^2\Pi^3_0 } \right]
\label{Q0}
\end{eqnarray}

\noindent Differentiating $Q_0$ with respect to $a$ yields

\begin{eqnarray}
{dQ_0 \over da} =  \sum^{\infty}_{m=1} \int^{\infty}_{\epsilon}{2x[x^2+ (2\pi m)^2] dx \over (x^2+(2\pi m)^2+0.125a [ x^2 + (2\pi m)^2 ]^{1/2} )^2} \nonumber \\
\times \left[ {1 \over x^2+[2\pi(m+1/2)]^2}+{1 \over x^2+[2\pi(m-1/2)]^2} \right] \nonumber \\
+\int^{\infty}_{\epsilon} {x^3 dx \over x^2 + \pi^2 } \left[ {1 \over [x^2+\beta^2 \Pi^1_0 ]^2 } + { 1 \over [ x^2+ \beta^2 \Pi^3_0 ]^2 } \right]
\end{eqnarray}

\noindent It is now obvious that, since $x \geq \epsilon \geq 0$, $\frac{dQ_0}{da} > 0$ and consequently $\frac{dQ_0}{dT}= -\frac{\alpha}{T^2} \ \frac{dQ_0}{da} < 0$ for any $T \neq 0$.
Noticing that the quantities $\beta^2 \Pi_{0}^{1}$, $\beta^2 \Pi_{0}^{3}$ are always positive for $x>0$ and $a>0$, we see that

\begin{eqnarray}
\frac{\partial Q(T,s^2)}{\partial s^2}=- a^3 \left\{ \sum_{m=1}^{\infty} \int_{\epsilon}^{\infty} \frac{2xdx}{x^2+(2\pi m)^2+0.125 a [x^2+(2\pi m)^2]^{1/2}} \right. \nonumber \\
\times \left[ \frac{1}{ [ x^2 + [ 2\pi (m + \frac{1}{2}) ]^2 + a^2 s^2]^2 } + \frac{1}{ [ x^2 + [ 2\pi (m - \frac{1}{2}) ]^2 + a^2 s^2]^2 } \right] \nonumber \\
 \left. + \int_{\epsilon}^{\infty} \frac{xdx}{[x^2 + \pi^2 + a^2s^2]^2} \left[ \frac{1}{x^2 +\beta^2 \Pi_{0}^{1}}+\frac{1}{x^2 +\beta^2 \Pi_{0}^{3}} \right] \right\}
\end{eqnarray}

\noindent is always negative. There are two more properties of $Q_{0}(T)$ which will come in use in the following, namely $\lim_{T \rightarrow 0} Q_{0}(T) = \infty $ which we discussed in section \ref{Numerical Results} and $\lim_{T \rightarrow \infty } Q_{0}(T) = 0$ which can be easily seen from (\ref{Q0}).
We are now ready to proceed to the discusssion of points (a) to (e).

(a) Consider the surface $Q(T,s^2)$. A critical temperature $T_c$ is a value of $T$ such that $s=0$ and at the same time the S-D equation is satisfied, that is $2 \pi N = Q_{0}(T) $. In geometrical terms it is the $T$-coordinate of a point which belongs both to the surface $Q(T,s^2)$ and to the line which is defined by the intersection of the planes $s^2=0$ and $Q=2 \pi N$. We have proved that the curve $Q_{0}(T)$ is decreasing with $T$ and $\lim_{T \rightarrow 0} Q_{0}(T) = \infty $, $\lim_{T \rightarrow \infty } Q_{0}(T) = 0$. Since $Q_{0}(T)$ is a continuous function of $T$, it follows that for any given $N$ there is a critical temperature $T_c=T_c(N)$. For any finite $N$, $T_c \neq 0$. The larger the $N$ is, the smaller the $T_c$ is in agreement with \cite{DoreyQED,DoreyThree}.

(b) For any $T > 0$, $Q_{0}(T)$ is monotonically decreasing with $T$. Thus the S-D equation $2 \pi N = Q_{0}(T)$ defines a unique $T_c$. 

(c) For any finite non-zero $T$ and any real non-zero $s$, $Q(T,s^2)<Q_{0}(T)$, because $Q(T,s^2)$ is a monotonically decreasing function of $s^2$. For $T>T_c$, $Q_{0}(T)<Q_{0}(T_c)=2 \pi N $, because $Q_{0}(T)$ is a monotonically decreasing function of $T$. Therefore $Q(T,s^2) < 2 \pi N$ which means that for $T>T_c$ there is no real non-zero $s$ which satisfies the S-D equations.
Combining (c) with (b), we see that, for $T>T_c$, only complex solutions for $s$ can exist. This would lead to unstable particles or tachyons. However above 
$T_c$ we can choose another solution to (\ref{sgapequation}), namely $s=0$ identically. Thus QED$_3$ reproduces the superconducting feature of the gap closing above some critical temperature.

(d) Consider that the curve $s=s(T)$ is a solution of the S-D equation, that is $2 \pi N = Q(T,s(T)^2)$. It is $s(T_c)=0$. In a neighboroughhood of $T_c$ we can make the approximation 

\begin{equation}
 Q(T,s(T)^2)=Q_{0}(T_c)+ \left[ R_T + R_S  \left. \frac{ds^2}{dT} \right|_{T_c}  \right] \ (T-T_c), 
\label{Taylor}
\end{equation}

\noindent where $ R_T= \left. \frac{\partial Q(T,s^2)}{\partial T} \right|_{T=T_c,s^2=0}<0 $ and $ R_s=  \left. \frac{\partial Q(T,s^2)}{\partial s^2} \right|_{T=T_c,s^2=0}<0 $. Recalling that $Q_{0}(T_c)=2 \pi N$, we can rewrite (\ref{Taylor}) as
\begin{equation}
2 s(T_c) \left. \frac{ds}{dT} \right|_{T_c} = - \frac{R_T}{R_S}  
\end{equation}

\noindent The right hand side of the above equation is a negative finite number for any finite temperature. Thus the fact that $s(T_c)=0^{+}$ implies that $\left. \frac{ds}{dT} \right|_{T_c} = - \infty$.

(e) We now turn our attention to arguing that $\left. {ds \over dT} \right|_{T=0}$ is cutoff independent. 
For this purpose we shall approximate $\Pi^1_0$ and $\Pi^3_0$ in two regions. In the first region we approximate both $\Pi^1_0$ and $\Pi^3_0$ by their 
low-$|{\bf k}|$ behaviour, namely $ \beta^2 \Pi^1_0 \simeq afx^2$ and $ \beta^2 \Pi^3_0 \simeq 2 \ln 2 a/\pi$; this region is to
 be defined by ${\bf k}$-momenta whose magnitude is below $\Lambda$. In the second region we approximate
 both $ \beta^2 \Pi^3_0$ and $ \beta^2 \Pi^1_0$ by $ax/8$, their behaviour
 at high-$|{\bf k}|$; we define this region by  $|{\bf k}|>\Lambda$. To make both $ \Pi^3_0 $ and $ \Pi^1_0$ continuous, $\Lambda$ should be chosen to be $16 ln2 / \pi$. In this argument we shall be considering $\epsilon<\Lambda$, although it should be straightforward to extend our analysis to $\epsilon>\Lambda$ (care being taken to include $\epsilon$ in other modes for $\epsilon>2 \pi$). Using these approximations we are able to write the contribution from the full zeroth mode as

\begin{equation}
{1 \over 2 \pi N} \left[ F_0(\epsilon,\Lambda,a,s)+ G_0(\Lambda,a,s) \right]
\end{equation}

\noindent where

\begin{equation}
F_0(\epsilon,\Lambda,a,s)=\int^{\Lambda}_{\epsilon} {ax \; dx \over x^2+a^2s^2+\pi^2} \left( {1 \over x^2+afx^2 }
+{1 \over x^2+2a \ln 2 /\pi} \right)
\end{equation} 

\noindent and

\begin{equation}
G_0(\Lambda,a,s)=2 \int^{\infty}_{\Lambda} {ax \; dx \over x^2+0.125ax} \left( {1 \over x^2+a^2s^2+\pi^2} \right).
\end{equation}

\noindent The integrals in these expressions can be evaluated analytically

\begin{eqnarray}
&&F_0(\epsilon,\Lambda,a,s)={a \over (1+af)(\pi^2+a^2s^2) } \left[ \ln \left( {\Lambda \over \epsilon}  \right)-{1 \over 2} \ln \left( {\Lambda^2+\pi^2+a^2s^2 \over \epsilon^2+\pi^2+a^2s^2 } \right) \right] 
\hspace{1cm} \nonumber \\
&& + \frac{a}{2 \left( \frac {2a \ln2}{\pi} - \pi^2 -a^2s^2 \right)} \left[ \ln \left( \frac{\Lambda^2+ 
\pi^2 + a^2 s^2}{\epsilon^2 + \pi^2 + a^2 s^2}  \right) - \ln \left( \frac{\Lambda^2+ \frac{2a \ln2}{\pi}} { \epsilon^2 + \frac{2a \ln2}{\pi}} \right) \right] \nonumber \\
&& G_{0}(\Lambda,a,s)={2a \over a^2s^2+\pi^2+(0.125)^2a^2} \left[ \ln \left( 
{ \sqrt{ \Lambda^2+a^2s^2+\pi^2} \over \Lambda+0.125a } \right) \right. \nonumber \\
&& \left. + {0.125a \over (a^2s^2+\pi^2)^{1/2}} \left[ {\pi \over 2} - \arctan \left( {\Lambda \over \sqrt{\pi^2+a^2s^2} } \right) \right] \right].  \hspace{1cm}
\end{eqnarray}

\noindent Since we are interested in the low $T$ behaviour of both $F_0$ and $G_0$, we expand up to second order in (1/a) for both terms, replacing leading order terms in $G_0$ by $I(0,0.125a,as)$. \
\begin{eqnarray}
&& F_0(\epsilon,\Lambda,a,s)  \simeq  {1 \over a^2} \tilde{F}(\epsilon,\Lambda,s) \equiv   \frac{1}{a^2} \left[ \frac{1}{fs^2} \ln \left( \frac{ \Lambda }{ \epsilon } \right) + \left( \Lambda^2 - \epsilon^2 \right) \frac{\pi}{4 s^2 \ln2} \right] \hspace{1cm} \\
&& G_0(\Lambda,a,s)  \simeq  {1 \over a^2} \tilde{G}(\Lambda,s) 
+2a I(0,0.125a,as) \equiv 
2a I(0,0.125a,as) - \nonumber \\ 
&& \frac{2}{a^2 (s^2+(0.125)^2)} \left( \frac{ \Lambda }{0.125}+ \frac{0.125 \Lambda}{s^2} \right) \hspace{1cm} 
\end{eqnarray}

\noindent We define $F_C$ as follows

\begin{equation}
\sum^{\infty}_{m=-\infty} 2a I(|2\pi m|,0.125a,(a^2 s^2+[2\pi(m+1/2))^2]^{1/2}) \equiv Q(\infty,s(T))-F_C(a,s)
\end{equation}

\noindent where $Q(\infty,s(T))$ coincides with the formula of section \ref{Numerical Results}, if $s(0)$ is replaced by $s(T)$, so that $F_C(\infty,s)=0$.
Now we can write (\ref{integralQED}) as

\begin{equation}
2 \pi N= {T^2 \over \alpha^2}[ \tilde{F}(\epsilon,\Lambda,s)+\tilde{G}(\Lambda,s)]+Q(\infty,s)-F_C(a,s).
\end{equation}

\noindent Differentiating the above equation with respect to $T$ yields an equation for ${dS \over dT}$, at low $T/\alpha$

\begin{eqnarray}
0={2T \over \alpha^2}[ \tilde{F}(\epsilon,\Lambda,s)+\tilde{G}(\Lambda,s)] 
+{T^2 \over \alpha^2} \left[  {\partial \tilde{F}(\epsilon,\Lambda,s) \over \partial s}+{\partial \tilde{G}(\Lambda,s) \over \partial s} \right]
 {ds \over dT} \nonumber \\ -F_{C}^{'}(a,s) + \frac{dQ(\infty,s(T))}{ds} \frac{ds}{dT}
\end{eqnarray}

\noindent where $F'_C={dF_C \over dT}$. Since $\partial \tilde{F} \over \partial s$, $\partial \tilde{G} \over \partial s$ are finite at $s>0$, $\epsilon>0$, this means that all the cutoff dependent terms vanish, for $s$ is non-singular at $T=0$ and we get a new gap equation which is cutoff independent. This means that $ \left. {ds \over dT} \right|_{T=0}$ is independent of $\epsilon$. To show that $ \left. {ds \over dT} \right|_{T=0}=0$ one must show that $F'_C(\infty,s)=0$ . Then one is left with the gap equation

\begin{equation}
0 =  \frac{ds}{dT} \ \frac{dQ(\infty,s)}{ds}
\end{equation}

\noindent for which the solution is $\frac{ds}{dT} =0$ for finite s.


\newpage

\renewcommand{\thefigure}{\arabic{figure}}
\setcounter{figure}{0}

\begin{figure}[p]
\begin{center}
\resizebox{\figwidth}{!}{\includegraphics*[134pt,435pt][465pt,625pt]{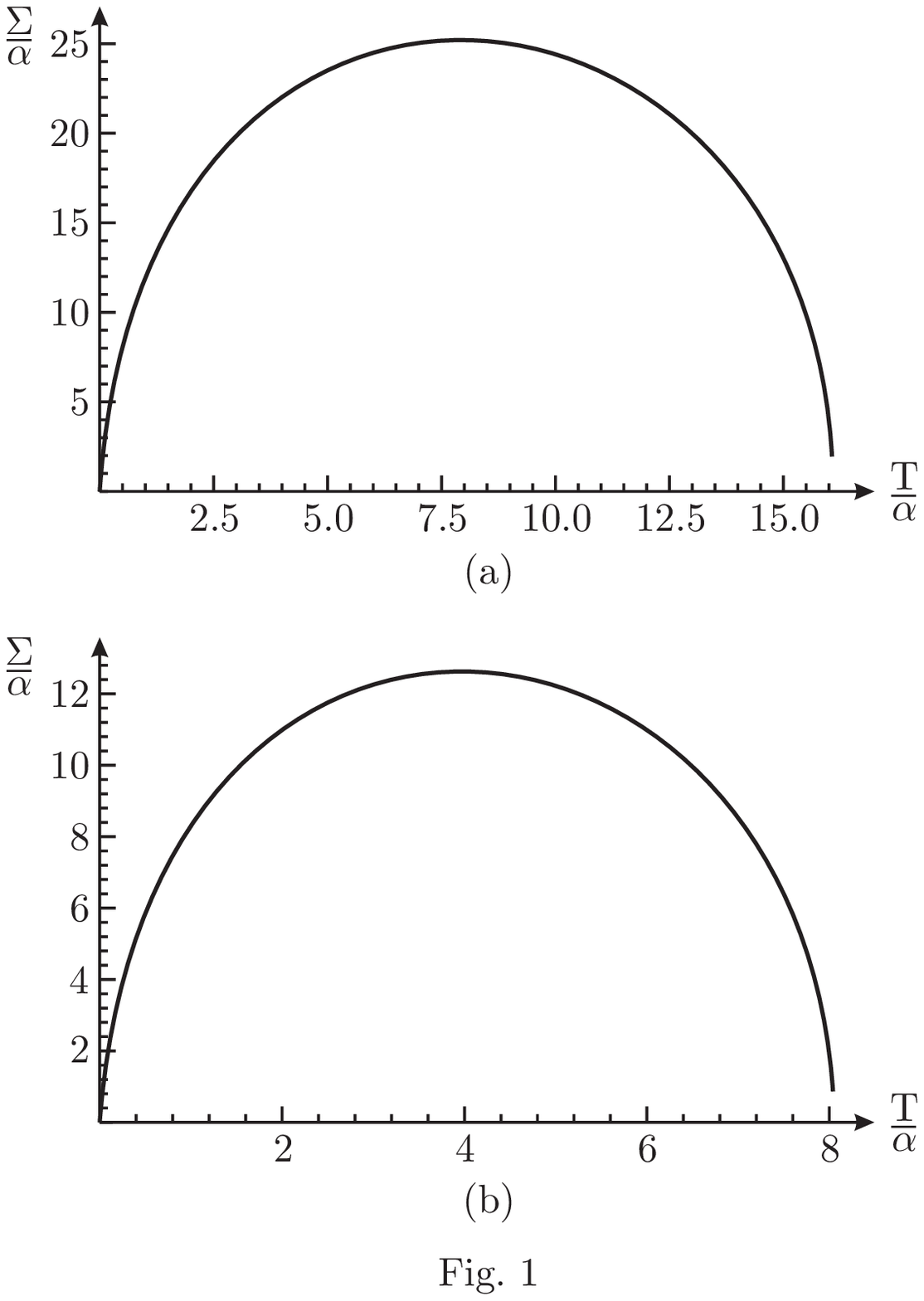}}
\caption{(a): The extreme solution for the integral cutoff case
($h=1000$).\label{fig.1aQED} }
\end{center}
\end{figure}

\setcounter{figure}{0}

\begin{figure}
\begin{center}
\resizebox{\figwidth}{!}{\includegraphics*[134pt,210pt][465pt,402pt]{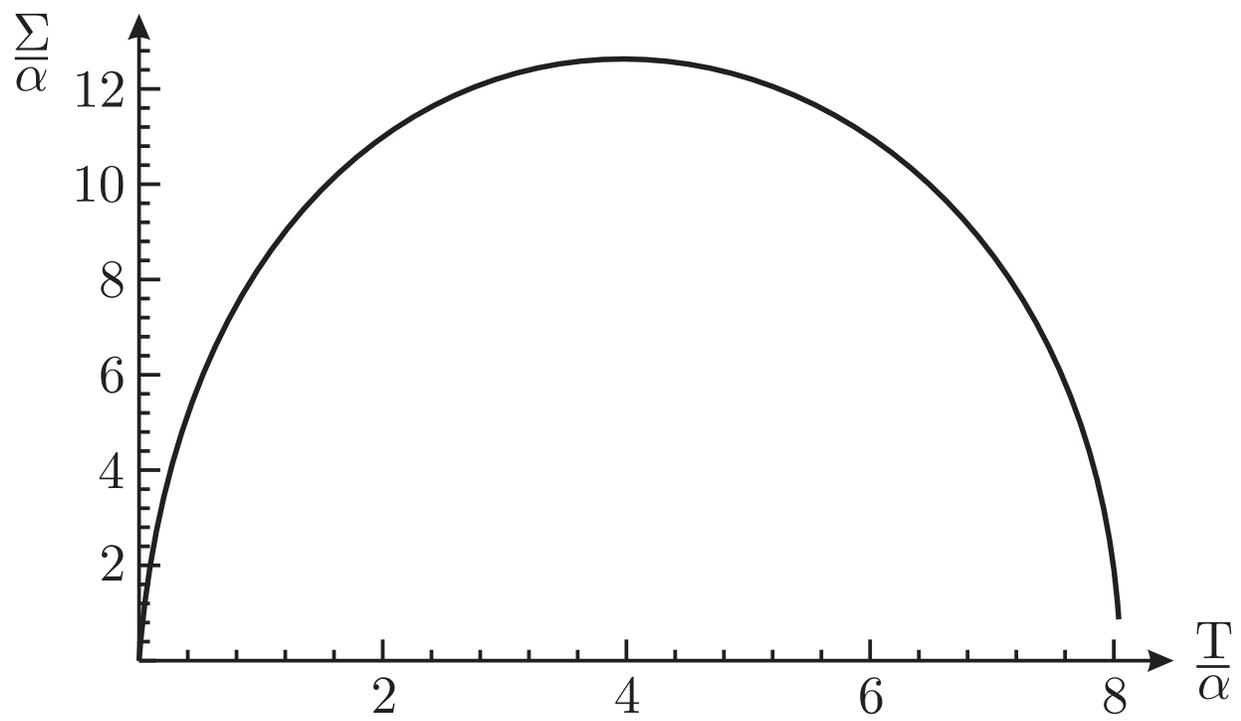}}
\caption{(b) The extreme solution for the mass cutoff case
($\tilde{h}=500$).\label{fig.1bQED}}
\end{center}
\end{figure}

\begin{figure}
\begin{center}
\resizebox{\figwidth}{!}{\includegraphics*[139pt,360pt][535pt,580pt]{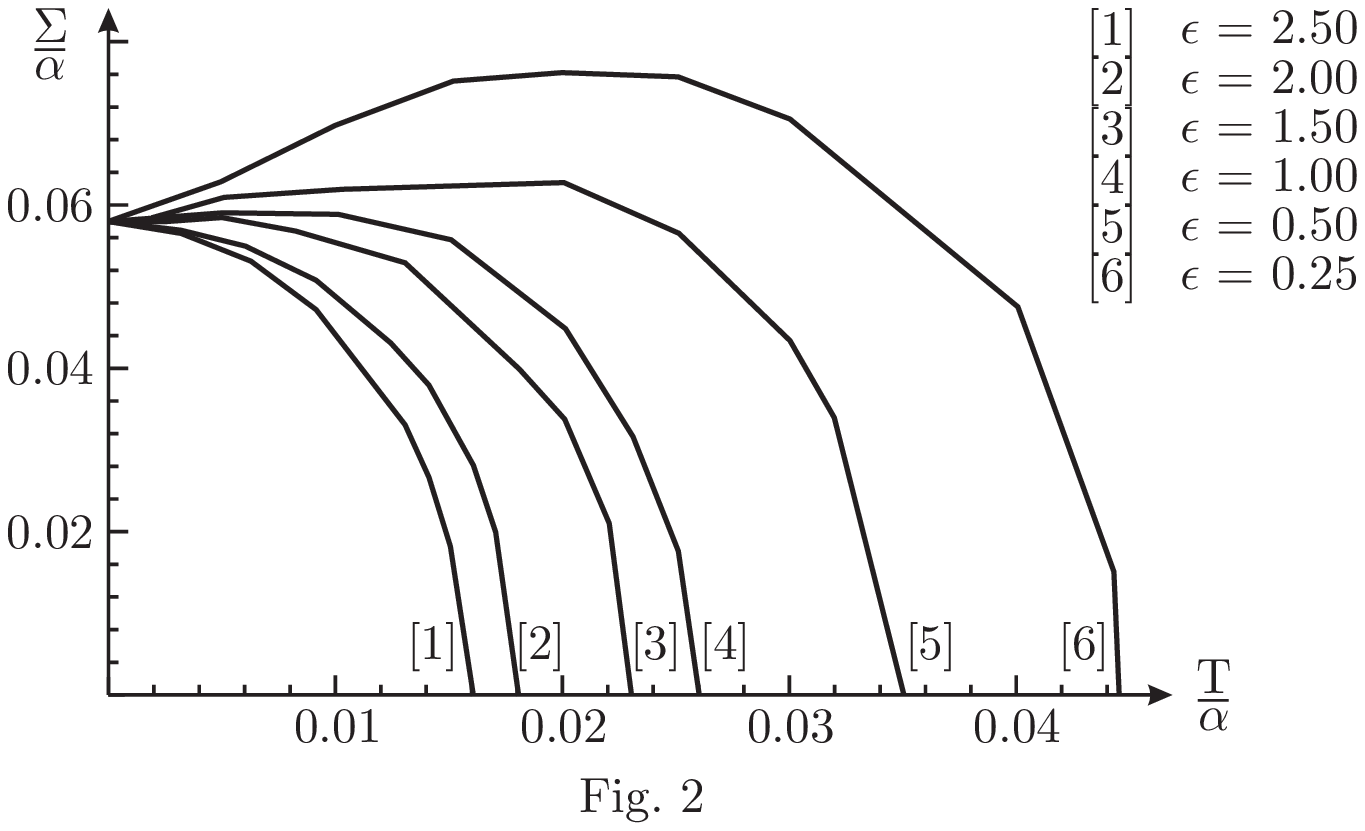}}
\caption{Numerical solutions for the integral cutoff $\epsilon$. The
curves show the scaled dynamically generated fermion mass as a
function of the scaled temperature for various values of $\epsilon$
and for $N=1$. \label{fig.2QED} }
\end{center}
\end{figure}

\begin{figure}
\begin{center}
\resizebox{\figwidth}{!}{\includegraphics*[109pt,340pt][535pt,620pt]{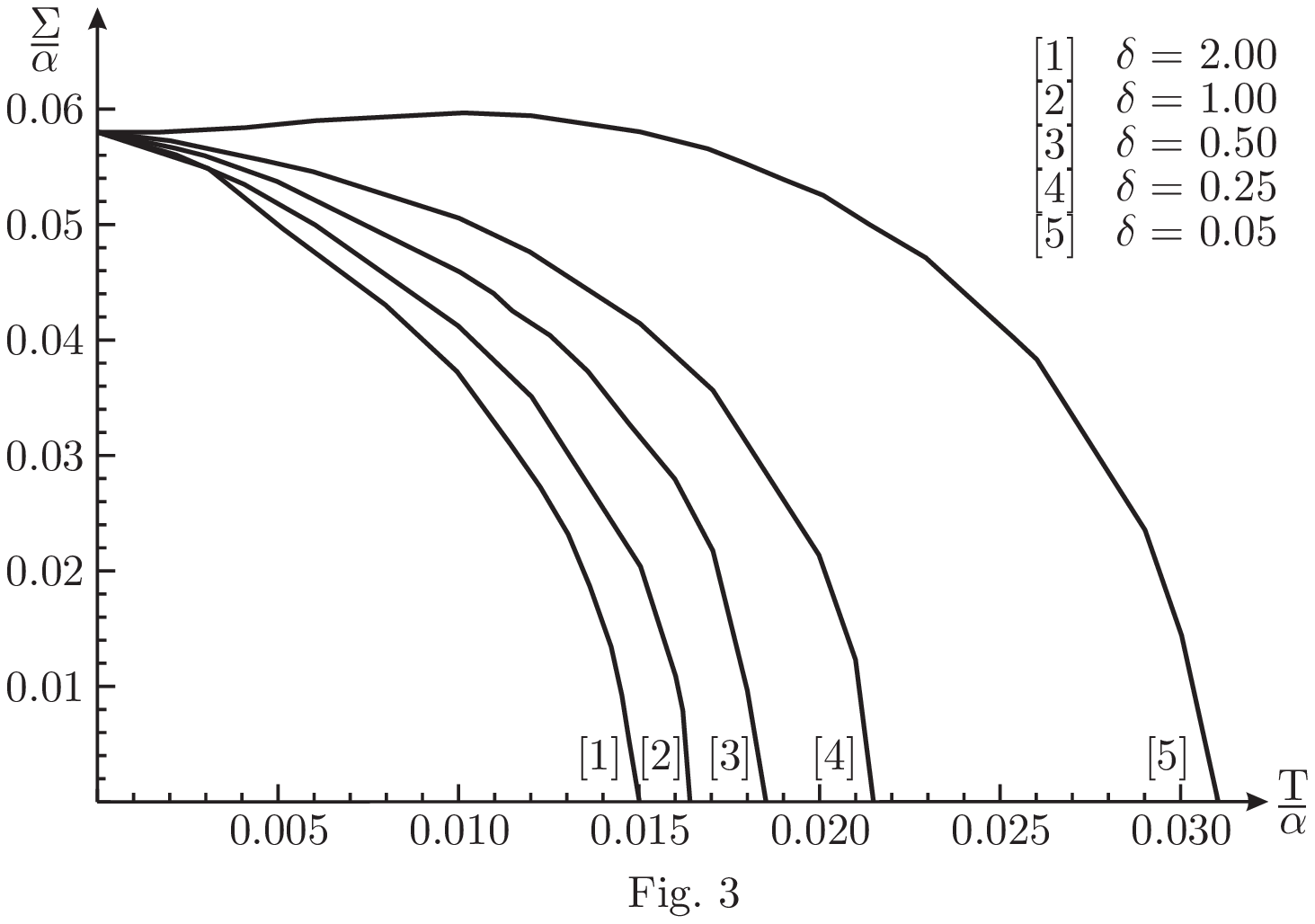}}
\caption{Numerical solutions for the mass cutoff $\delta$. The curves
 show the scaled dynamically generated fermion mass as a function of
 the scaled temperature for various values of $\delta$ and for $N=1$.\label{fig.3QED} }
\end{center}
\end{figure}

\begin{figure}
\begin{center}
\resizebox{\figwidth}{!}{\includegraphics*[109pt,280pt][570pt,595pt]{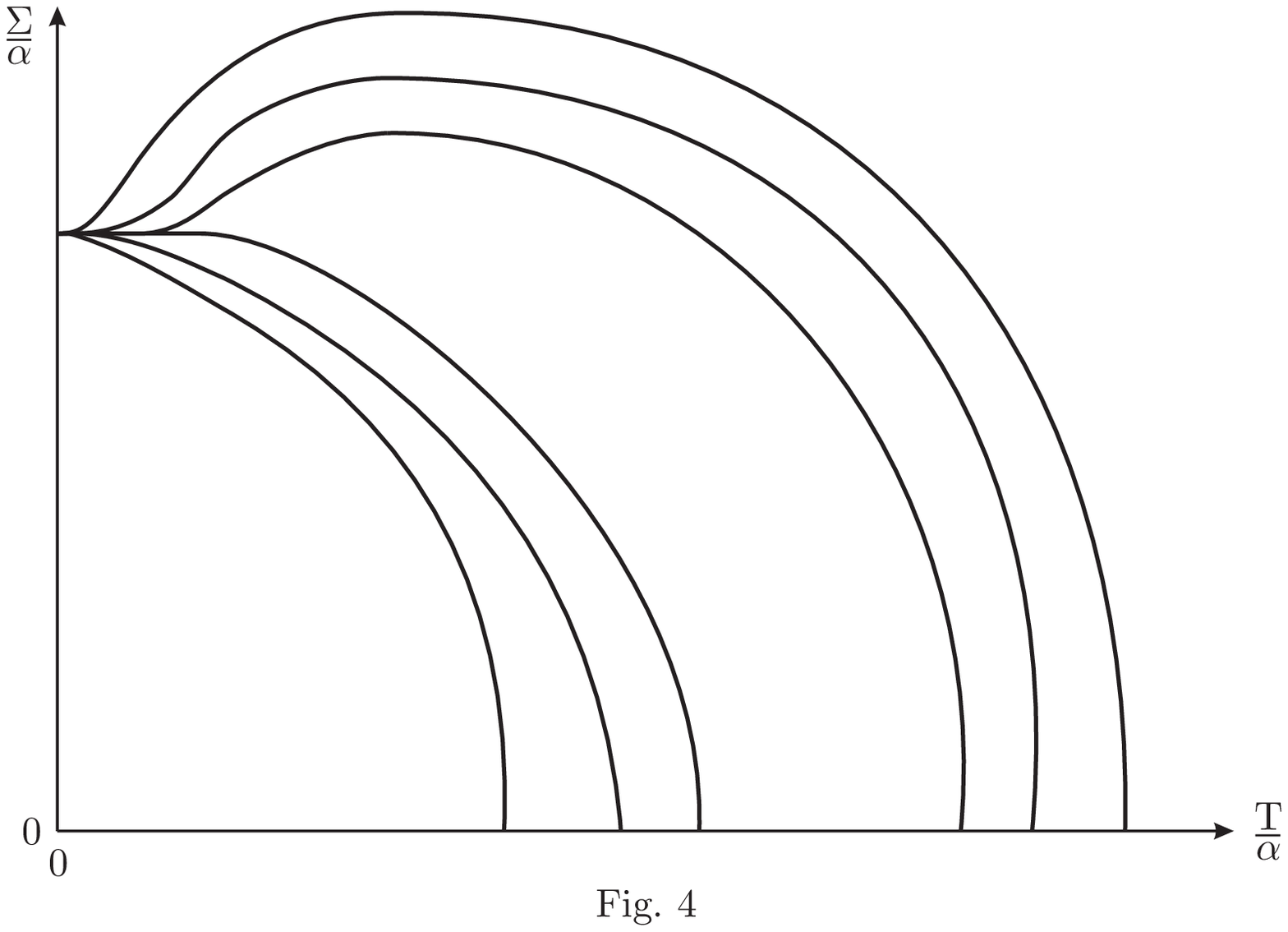}}
\caption{Each curve corresponds to a different value of $\epsilon$ (or
$\delta$). All the curves have zero gradient as $T \rightarrow 0$,
$\lim_{T\rightarrow 0} \frac{d\Sigma}{dT}=0$, satisfying thus the third
law of thermodynamics.\label{fig.4QED}}
\end{center}
\end{figure}

\begin{figure}
\begin{center}
\resizebox{\figwidth}{!}{\includegraphics*[88pt,490pt][515pt,660pt]{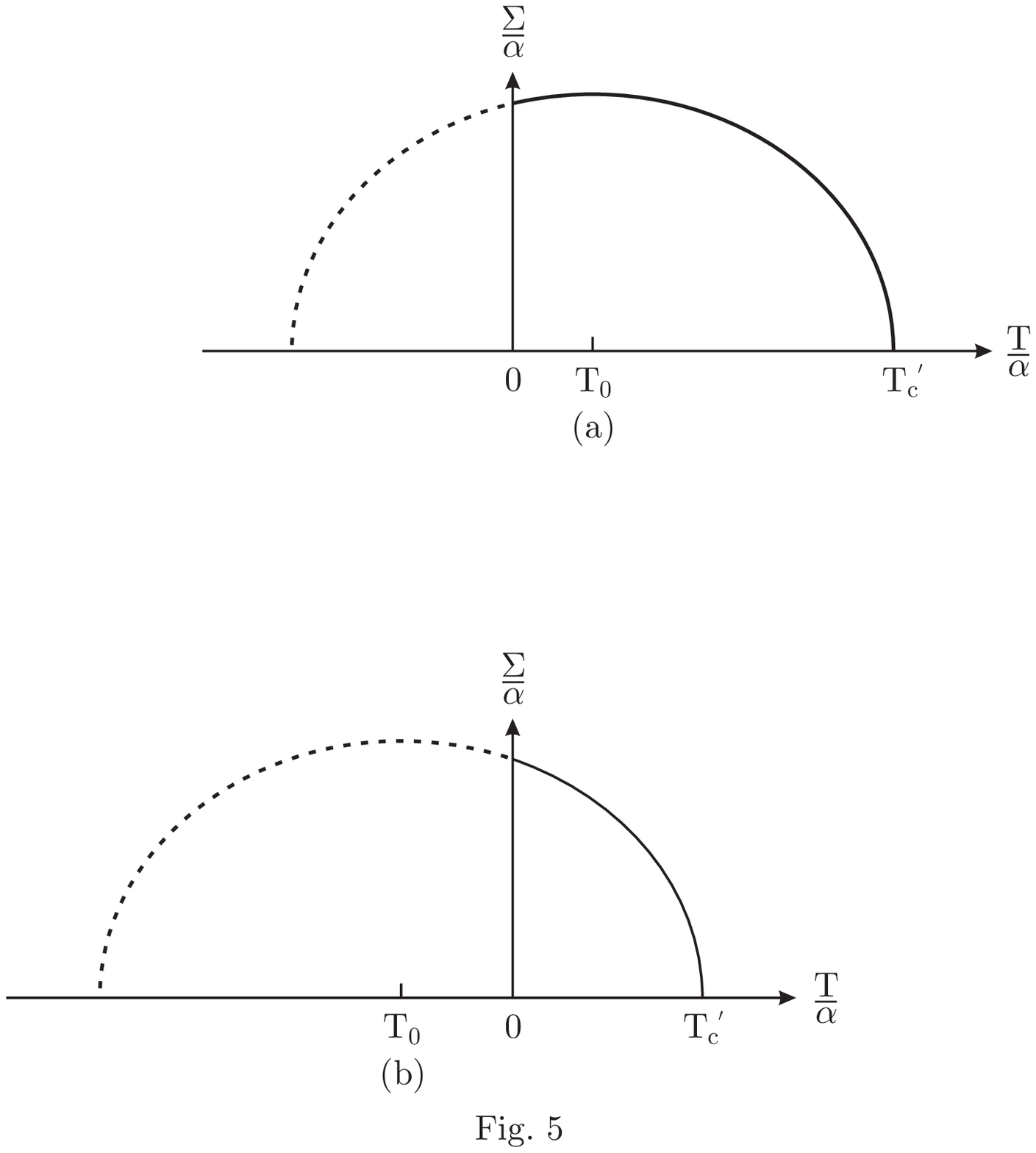}}
\caption{(a): The ellipse with $T_{0}>0$. $\frac{d\Sigma}{dT} >0$ for
$\Sigma > 0$ and $0<T<T_{0}$, $\frac{d\Sigma}{dT}<0$ for $\Sigma >0$
and $T_{0}<T<T^{\prime}_{c}$. \label{fig.5aQED}}
\end{center}
\end{figure}

\setcounter{figure}{4}

\begin{figure}
\begin{center}
\resizebox{\figwidth}{!}{\includegraphics*[88pt,227pt][515pt,433pt]{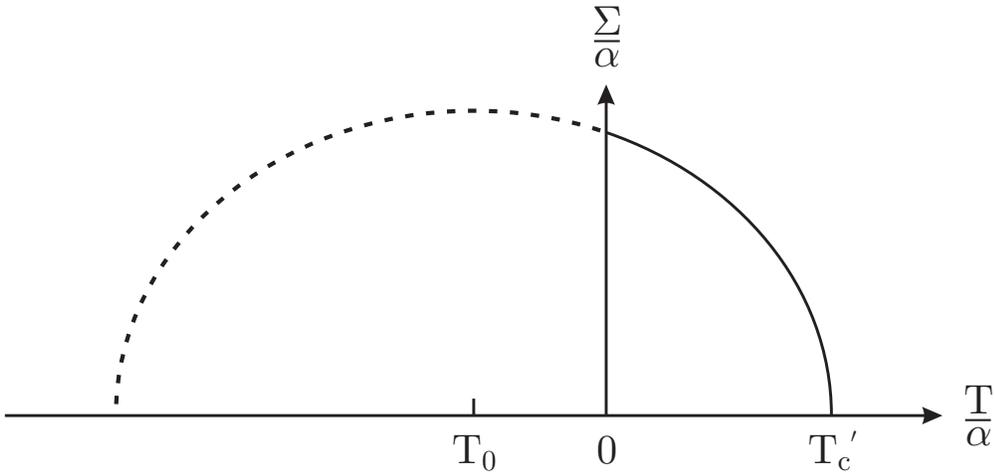}}
\caption{(b): The ellipse with
$T_{0}<0$. $\frac{d\Sigma}{dT}<0$ for $\Sigma >0$ and
$0<T<T^{\prime}_{c}$.\label{fig.5bQED}}
\end{center}
\end{figure}

\begin{figure}
\begin{center}
\resizebox{\figwidth}{!}{\includegraphics*[50pt,283pt][570pt,602pt]{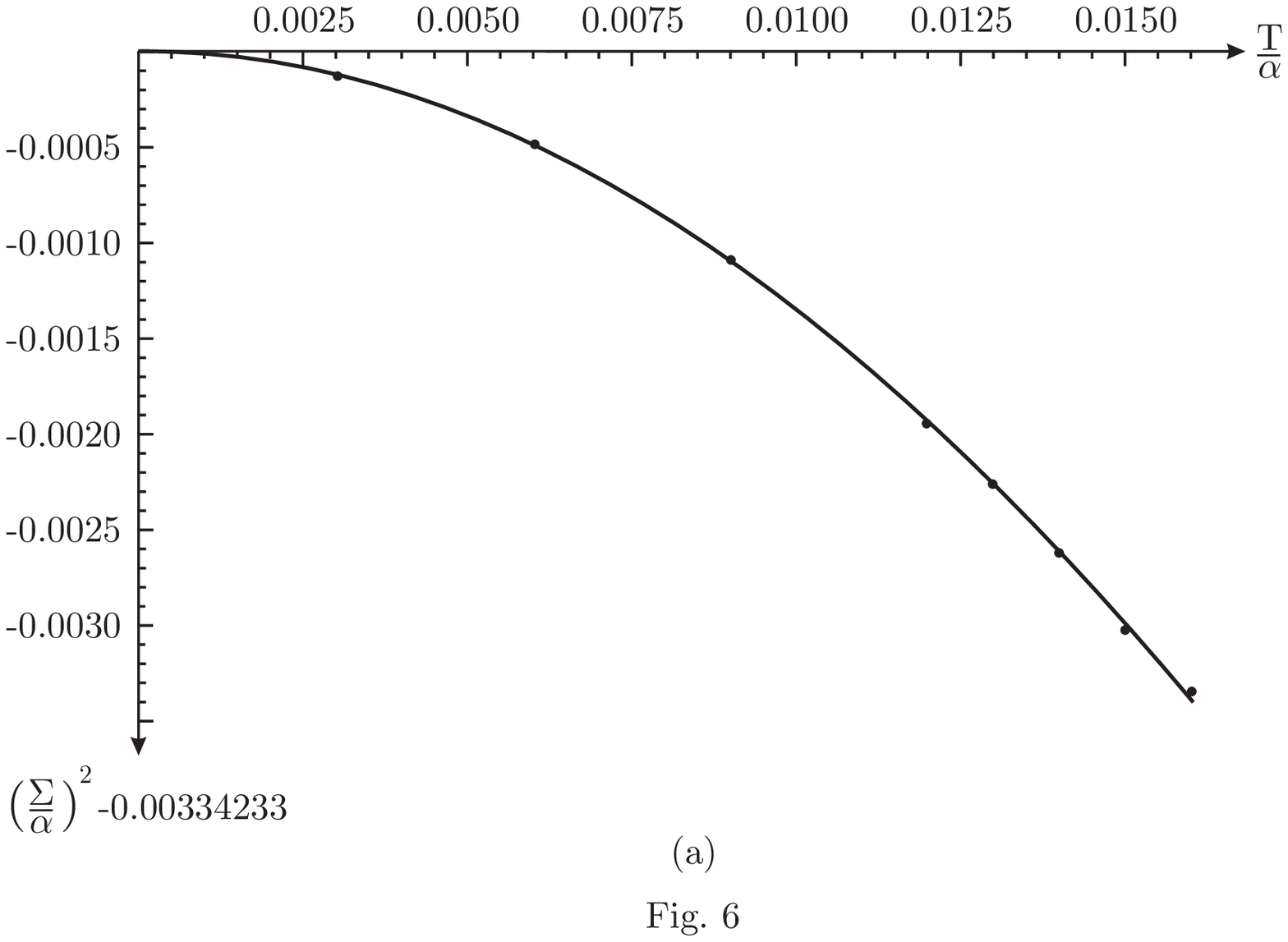}}
\caption{(a): The diagram shows the fitting of the ellipse approximation
$ \left( \frac{ \Sigma }{ \alpha } \right)^2 = 0.00334233 - 0.00546513
\left( \frac{T}{ \alpha } \right) - 12.8907 \left( \frac{T}{ \alpha }
\right)^2 $ to the points we have found for $ \epsilon = 2.5 $. 
\label{fig.6aQED} } 
\end{center}
\end{figure}

\setcounter{figure}{5}

\begin{figure}
\begin{center}
\resizebox{\figwidth}{!}{\includegraphics*[50pt,283pt][570pt,602pt]{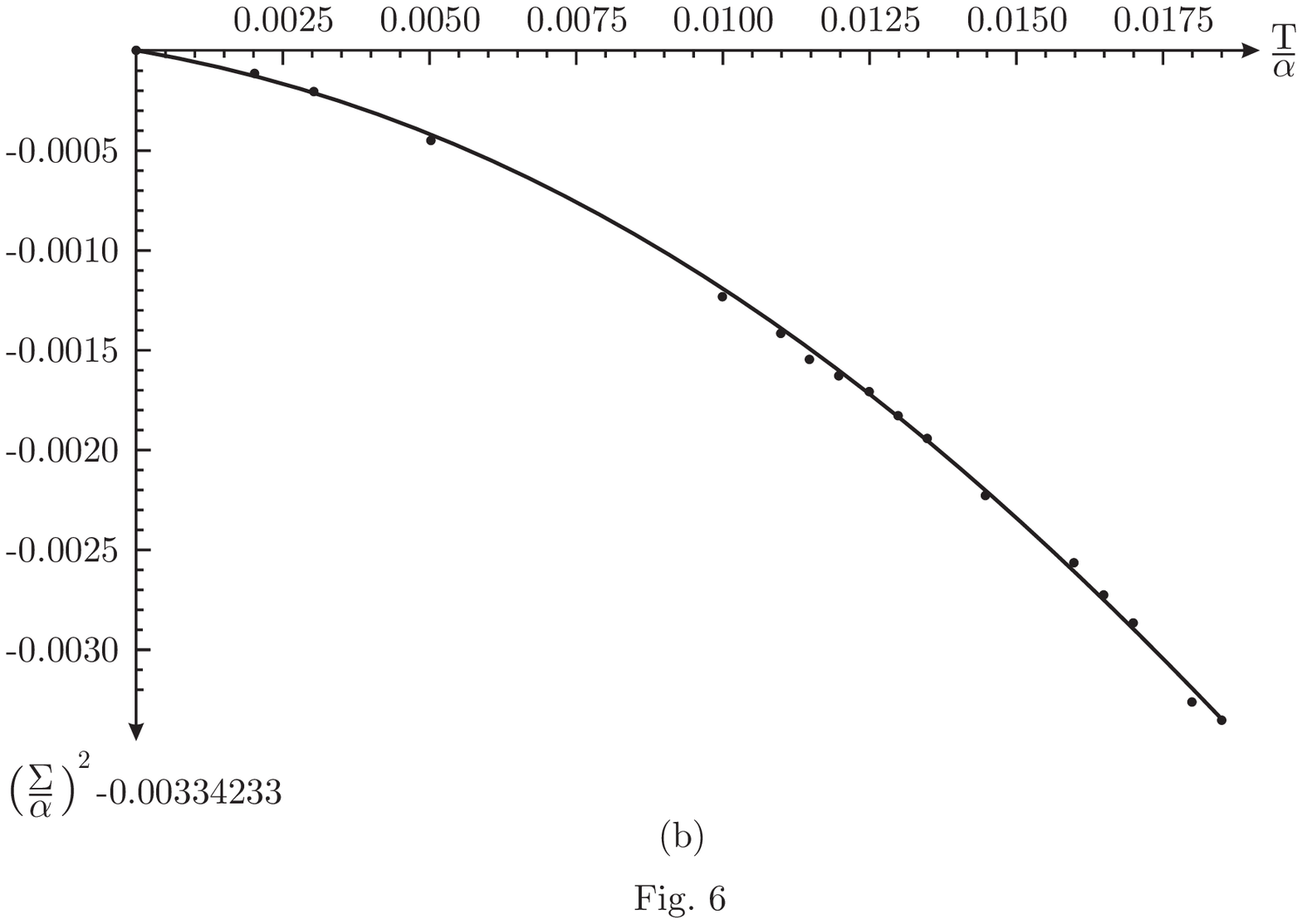}}
\caption{(b): The diagram shows the fitting of the ellipse approximation 
$ \left( \frac{ \Sigma }{ \alpha } \right)^2 = 0.00334233 - 0.0497448
\left( \frac{ T  }{ \alpha } \right) -7.05771 \left( \frac{ T }{
\alpha } \right)^2 $ to the points we have found for $ \delta = 0.5 $.
\label{fig.6bQED}}
\end{center}
\end{figure}

\begin{figure}
\begin{center}
\resizebox{\figwidth}{!}{\includegraphics*[50pt,295pt][570pt,602pt]{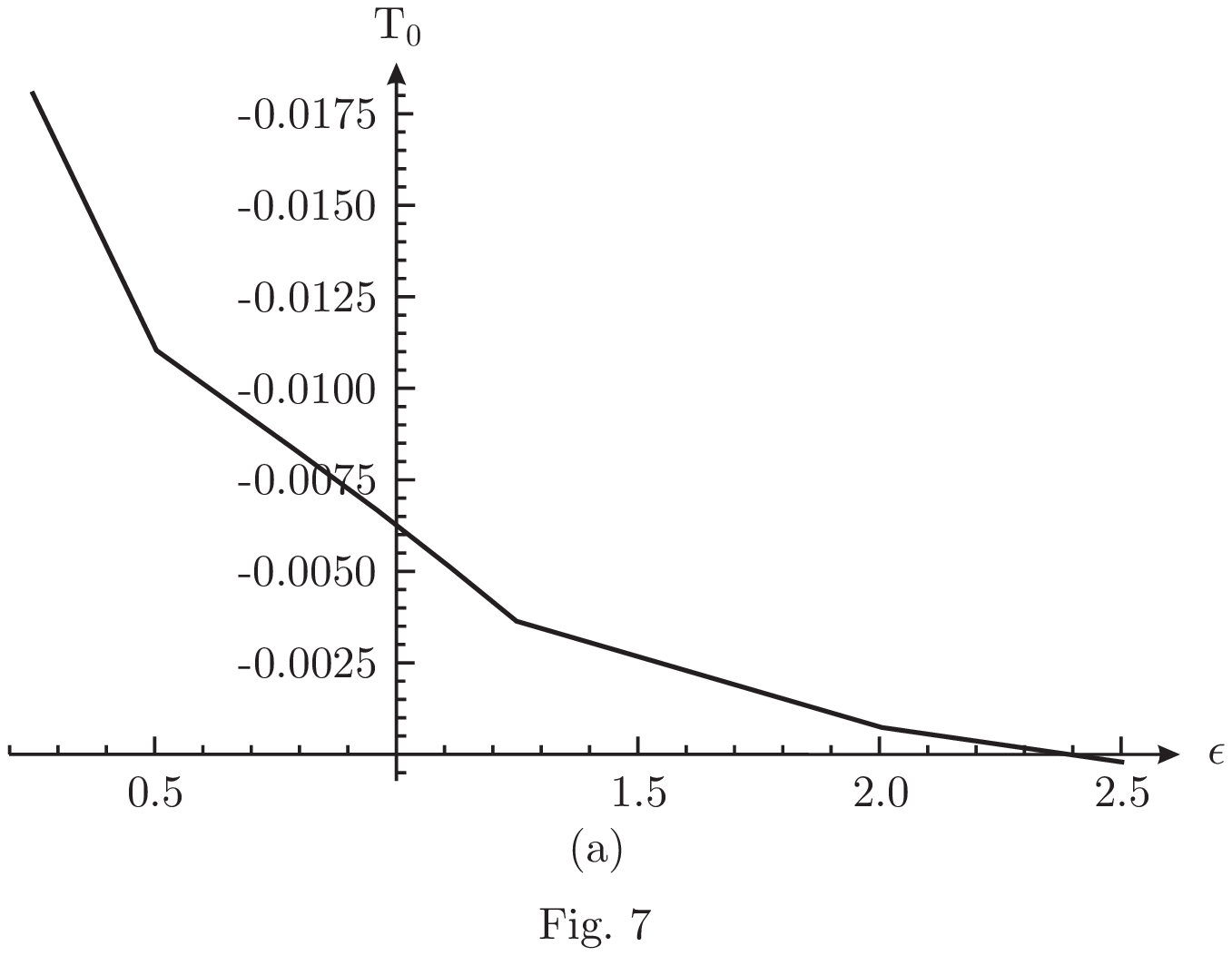}}
\caption{(a): The centre of the ellipse $ T_0 $ as a function of the integral cutoff  $ \epsilon $. $ T_0 = 0 $ at $ \epsilon_c \simeq 2.4 $.\label{fig.7aQED} }
\end{center}
\end{figure}

\setcounter{figure}{6}

\begin{figure}
\begin{center}
\resizebox{\figwidth}{!}{\includegraphics*[50pt,352pt][570pt,602pt]{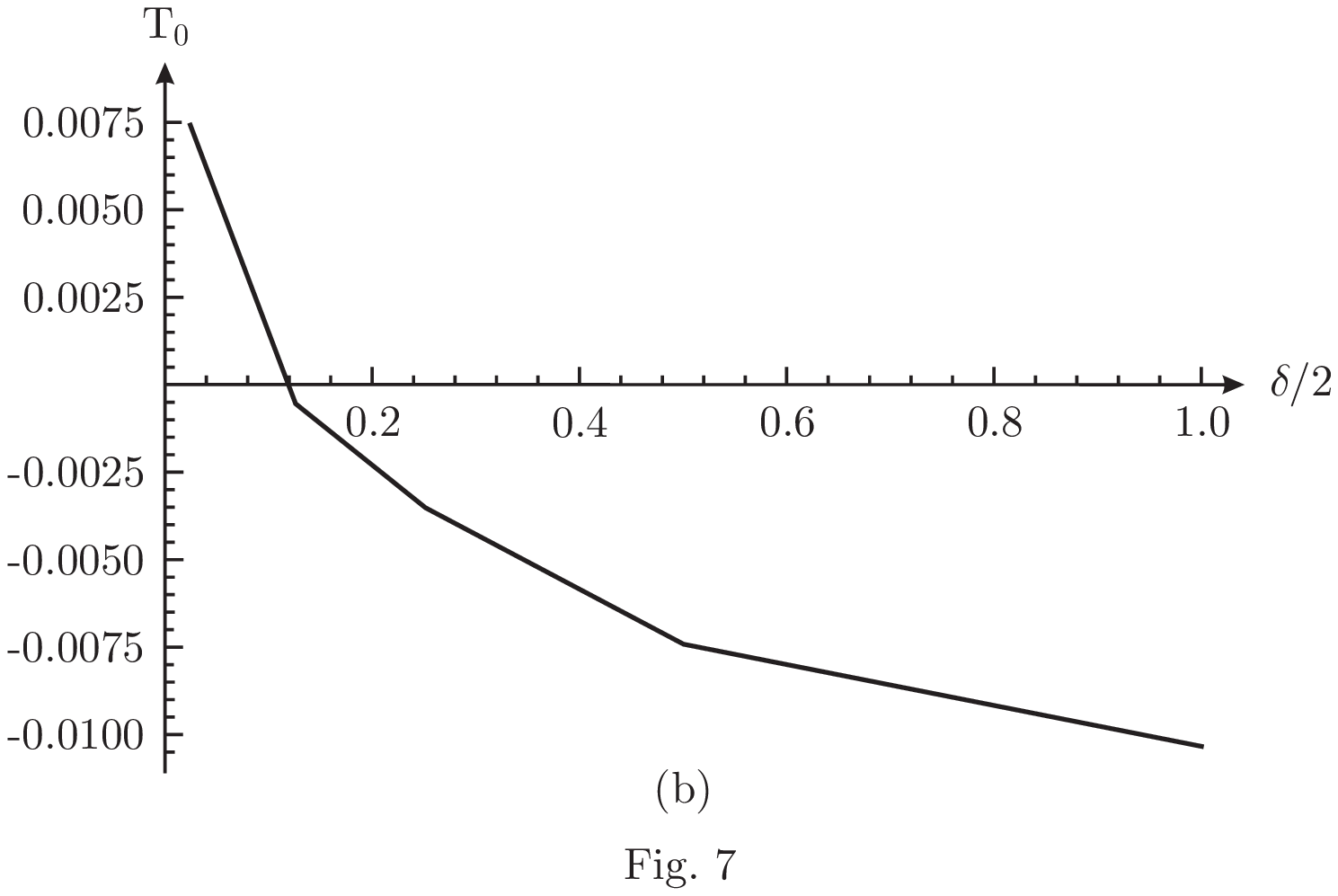}}
\caption{(b): The centre of the ellipse $ T_0 $ as a function of the mass cutoff $ \delta $. $T_0 = 0$ at $ \delta \simeq 0.22 $.\label{fig.7bQED}}
\end{center}
\end{figure}

\end{document}